\documentclass[sigconf,nonacm]{acmart}

\usepackage{acmart-taps}

\usepackage{tabularray}
\UseTblrLibrary{booktabs}
\usepackage{multirow} % Combine rows in tables
\usepackage{array} % Extended array and tabular environments
\newcolumntype{C}[1]{>{\centering\arraybackslash}p{#1}} % Centered version of p-column

\usepackage{wrapfig} % Wrap text around figures
\usepackage{float} % Improved control over float positioning

\usepackage{siunitx} % Consistent typesetting of units and numbers
\usepackage{etoolbox} % Conditional and robust commands
\usepackage{dcolumn} % Align decimal columns in tables
\usepackage{booktabs} % Professional-looking tables (rules, spacing)

\usepackage{fontawesome5}
\usepackage{svg}
\usepackage{xcolor} % Provides color support
\usepackage{soul} % Highlighting, underlining, and other text decorations
\soulregister{\cite}7 % Register cite commands so they won't break under soul
\soulregister{\citep}7
\soulregister{\citet}7
\soulregister{\ref}7
\soulregister{\pageref}7

\usepackage[most]{tcolorbox}

\newtcolorbox{myfancybox}{
  colback=light-bg,
  colframe=black,
  boxrule=0.8pt,
  arc=0pt,
  outer arc=0pt,
  left=6pt,
  right=6pt,
  top=6pt,
  bottom=6pt,
  before skip=8pt,
  after skip=8pt,
  parskip=3pt
}

\usepackage{xspace} % Insert space intelligently after macros

\usepackage{bm} % Bold math symbols
\newrobustcmd*{\bftabnum}{ %
  \bfseries
  \sisetup{output-decimal-marker={\textmd{.}}} %
}

\usepackage{hyperref}
\usepackage{cleveref}

\usepackage{tikz}
\usetikzlibrary{arrows.meta,positioning,fit,backgrounds,calc}
\definecolor{oxfordblue}{rgb}{0.0, 0.13, 0.28}
\definecolor{harvardcrimson}{rgb}{0.79, 0.0, 0.09}
\definecolor{dartmouthgreen}{rgb}{0.05, 0.5, 0.06}
\definecolor{princetonorange}{rgb}{1.0, 0.56, 0.0}
\definecolor{yaleblue}{rgb}{0.06, 0.3, 0.57}
\definecolor{usccardinal}{rgb}{0.6, 0.0, 0.0}
\definecolor{uclablue}{rgb}{0.33, 0.41, 0.58}
\definecolor{msugreen}{rgb}{0.09, 0.27, 0.23}
\definecolor{cornellred}{rgb}{0.7, 0.11, 0.11}
\definecolor{pomegranate}{RGB}{192, 57, 43}
\definecolor{anti-pomegranate}{RGB}{43,178,192}
\definecolor{alizarin}{RGB}{231, 76, 60}
\definecolor{anti-belize}{RGB}{185, 41, 56}
\definecolor{belize}{RGB}{41, 128, 185}
\definecolor{sky}{RGB}{52, 152, 219}
\definecolor{green}{RGB}{22, 160, 133}
\definecolor{anti-green}{RGB}{160,22,118}
\definecolor{turquoise}{RGB}{26, 188, 156}
\definecolor{pumpkin}{RGB}{211, 84, 0}
\definecolor{anti-pumpkin}{RGB}{0,22,211}
\definecolor{carrot}{RGB}{230, 126, 34}
\definecolor{wisteria}{RGB}{142, 68, 173}
\definecolor{anti-wisteria}{RGB}{99,173,68}
\definecolor{amethyst}{RGB}{155, 89, 182}
\definecolor{nephritis}{RGB}{39, 174, 96}
\definecolor{anti-nephritis}{RGB}{174,39,117}
\definecolor{grey-bg}{RGB}{242,242,235}
\definecolor{light-bg}{RGB}{249,249,249}
\definecolor{extended-blue}{RGB}{59,130,246}
\definecolor{extended-red}{RGB}{239,68,68}
\definecolor{extended-orange}{RGB}{249,115,22}
\definecolor{extended-violet}{RGB}{99,102,241}
\definecolor{extended-green}{RGB}{16,185,129}

\newcommand{\remove}[1]{}
\newcommand{\revision}[1]{{#1}}
\newcommand{\eg}{e.g.,\ }

\newcommand{\ie}{i.e.,\ }

\newcommand{\tool}{\textsc{Malleable Prompting}\xspace}

\AtBeginDocument{ %
  \providecommand\BibTeX{{ %
    \normalfont B\kern-0.5em{\scshape i\kern-0.25em b}\kern-0.8em\TeX}}}

\newcommand{\namedparagraph}[1]{\vspace{0.2cm}\noindent\textbf{#1:}}

\newcommand{\formatcaption}[2]{\textit{#1} \textmd{#2}}

\newcommand{\fig}[1]{Fig. {#1}}
\newcommand{\figref}[1]{\fig{\ref{#1}}}
\newcommand{\tab}[1]{Table {#1}}
\newcommand{\tabref}[1]{\tab{\ref{#1}}}
\newcommand{\secref}[1]{§\ref{#1}}

\usepackage{enumitem}
\newenvironment{packeditemize}{
\begin{itemize}[leftmargin=0.5cm]
\setlength{\itemsep}{1pt}
\setlength{\parskip}{2pt}
\setlength{\parsep}{0pt}
}{\end{itemize}}

\newenvironment{packedenumerate}{
\begin{enumerate}[leftmargin=0.5cm]
\setlength{\itemsep}{1pt}
\setlength{\parskip}{2pt}
\setlength{\parsep}{0pt}
}{\end{enumerate}}

\usepackage{pifont}  % For xmark

\newcommand{\xmark}{\textcolor{black}{\ding{55}}}% Xmark

\usepackage{xcolor}

\definecolor{neonfuchsia}{rgb}{1.0, 0.25, 0.39}

\newcommand{\userquote}[1]{\textit{``#1''}}

\newcommand{\influence}[1]{\textcolor{cornellred}{#1}}

\newcommand{\maxnumber}[1]{{\setlength{\fboxsep}{1pt}\colorbox{yellow!25}{#1}}}

\acmSubmissionID{1266}

\copyrightyear{2026}
\acmYear{2026}
\setcopyright{cc}
\setcctype{by}
\acmConference[UIST '26]{The 39th Annual ACM Symposium on User Interface Software and Technology}{November 02--05, 2026}{Detroit, MI, USA}
\acmBooktitle{The 39th Annual ACM Symposium on User Interface Software and Technology (UIST '26), November 02--05, 2026, Detroit, MI, USA}
\acmDOI{10.1145/3830398.3830587}
\acmISBN{979-8-4007-2856-3/2026/11}

\begin{document}

\title[Malleable Prompting]{From Words to Widgets for Controllable LLM Generation}

\author{Chao Zhang}
\authornote{These authors contributed equally to this work.}
\email{cz468@cornell.edu}
\orcid{0000-0003-4286-8468}
\affiliation{ %
 \institution{Cornell University}
 \city{Ithaca, NY}
 \country{USA}}

\author{Yiren Liu}
\authornotemark[1]
\email{yirenl2@illinois.edu}
\orcid{0000-0003-1507-0303}
\affiliation{%
 \institution{University of Illinois Urbana-Champaign}
 \city{Champaign, IL}
 \country{USA}}

\author{Lunyiu Nie}
\authornotemark[1]
\email{lynie@utexas.edu}
\orcid{0000-0002-6721-4578}
\affiliation{%
 \institution{The University of Texas at Austin}
 \city{Austin, TX}
 \country{USA}}

\author{Jeffrey M. Rzeszotarski}
\email{jrzeszotarski@loyola.edu}
\orcid{0000-0002-4317-9501}
\affiliation{%
 \institution{Loyola University Maryland}
 \city{Baltimore, MD}
 \country{USA}}

\author{Yun Huang}
\email{yunhuang@illinois.edu}
\orcid{0000-0003-0399-8032}
\affiliation{%
 \institution{University of Illinois Urbana-Champaign}
 \city{Champaign, IL}
 \country{USA}}

\author{Tal August}
\email{taugust@illinois.edu}
\orcid{0000-0001-6726-4009}
\affiliation{%
 \institution{University of Illinois Urbana-Champaign}
 \city{Champaign, IL}
 \country{USA}}

\renewcommand{\shortauthors}{Zhang et al.}

\begin{abstract}
% Natural language remains the predominant way people interact with Large language models (LLMs), yet users often struggle to precisely control subjective preferences, such as tone, style, and emphasis, through back-and-forth prompt editing. 
Natural language remains the predominant way people interact with large language models (LLMs).
However, users often struggle to precisely express and control subjective preferences (\eg tone, style, and emphasis) through prompting. 
We propose \tool, a new interactive prompting technique for controllable LLM generation.
% We observe that conversational prompting often breaks down when users refine subjective preferences (e.g., tone, concision, emphasis): it is hard to precisely articulate intent in natural language, and it is hard to compare drafts or attribute changes as instructions accumulate across turns. 
It reifies preference expressions in natural language prompts into GUI widgets (\eg sliders, dropdowns, and toggles) that users can directly configure to steer generation, while visualizing each control's influence on the output to support attribution and comparison across iterations. 
To enable this interaction, we introduce an LLM decoding algorithm that modulates the token probability distribution during generation based on preference expressions and their widget values.
% To enable \tool, we introduce a decoding-time token influence algorithm that isolates the effect of each preference expression on the token distribution at inference time, enabling independent control and attribution.
% Through a user study, we show that \tool helps participants achieve target preferences more precisely and perceive prompting editing using \tool as more controllable and transparent.
Through a user study, we show that \tool helps participants achieve target preferences more precisely and is perceived as more controllable and transparent than natural language prompting alone.
% Our findings suggest that externalizing preferences as parameters enables more predictable and controllable prompting interaction.

% \vspace{12\baselineskip}

\end{abstract}

% CCS XML, Keywords, etc.



\begin{teaserfigure}
 \includegraphics[width=\linewidth]{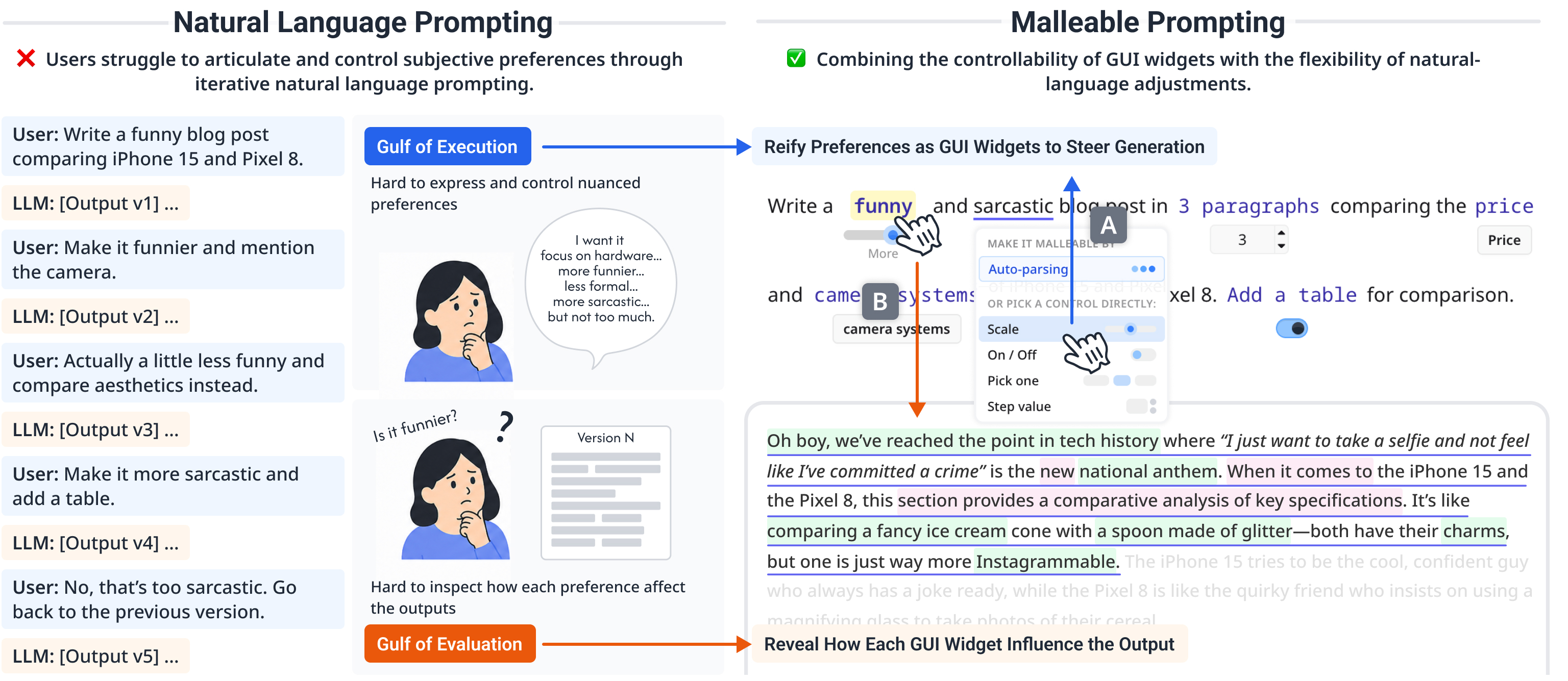}
  \caption{\formatcaption{\tool.}{\revision{Instead of repeatedly refining outputs through underspecified natural-language prompts (left), \tool lets users reify prompt preferences as GUI widgets and directly manipulate them to steer generation (A). Hovering over a widget reveals the output spans influenced by that preference, making individual effects inspectable (B).}}}
 \Description{A two-part teaser figure contrasts natural-language prompting with malleable prompting. On the left, titled ``Natural Language Prompting,'' a sequence of chat messages shows a user repeatedly revising a request for a funny blog post comparing the iPhone 15 and Pixel 8: asking to make it funnier, mention the camera, reduce humor, compare aesthetics, add sarcasm, add a table, and then revert an overly sarcastic result. Two illustrations identify the resulting interaction problems. The ``Gulf of Execution'' shows a user thinking through several nuanced and partially conflicting preferences, such as focusing on hardware, being funnier but less formal, and adding only a limited amount of sarcasm; the caption states that these preferences are hard to express and control. The ``Gulf of Evaluation'' shows the user looking uncertainly at a document version; the caption states that it is hard to understand how each preference affects the output. On the right, titled ``Malleable Prompting,'' the figure presents Malleable prompting as combining GUI-widget controllability with the flexibility of natural-language adjustments. A prompt contains visually emphasized preference phrases, including ``funny,'' ``3 paragraphs,'' ``price,'' ``camera systems,'' and ``Add a table.'' A contextual widget panel converts these phrases into appropriate controls, including a scale slider, on/off toggle, pick-one selector, and step-value control. A blue annotation states, ``Reify Preferences as GUI Widgets to Steer Generation.'' Below, the generated blog-post text contains colored underlines and highlights corresponding to different preference widgets. An orange annotation states, ``Reveal How Each GUI Widget Influence the Output,'' indicating that hovering over a widget exposes the text spans it affects.}
 \label{fig:teaser}
\end{teaserfigure}

\maketitle

\section{Introduction}
\label{sec:intro}
% Prompts are the vehicle for humans to express \textit{preferences}---the fuzzy, subjective qualities of text, such as tone, style, and emphasis---for large language model (LLM) output.
% These preferences act as ambiguous and multifaceted constraints that users painstakingly tweak to steer LLMs. 
When interacting with large language models (LLMs), users provide instructions via prompts, yet models rarely generate content that aligns perfectly with user intent on the first try.
As a result, users often engage in an iterative process~\cite{mysorePrototypicalHumanAICollaboration2025a}, refining the model's output through turn-by-turn conversations to better match their preferences.
For example, a user may ask a model to draft a blog post comparing two phones, then iterate by specifying comparison aspects (\eg price, battery, aesthetics), altering the structure (bulleted vs. narrative), and fine-tuning style (\eg ``more concise,'' ``less formal'').
% While prior work has helped users revise concrete aspects of generated text, such as replacing words with synonyms, through direct manipulation~\cite{masson_directgpt_2024}---for instance, by selecting spans in the output---users still struggle to express and control subjective \textit{preferences}, such as tone, style, and emphasis.
While prior work has helped users revise concrete aspects of generated text, such as replacing words with synonyms, through direct manipulation~\cite{masson_directgpt_2024}, users still struggle to articulate and control more subjective \textit{preferences}---such as tone, style, and emphasis---through prompting alone~\cite{liElicitingHumanPreferences2023,subramonyamBridgingGulfEnvisioning2024}.
This challenge stems from two interaction gulfs in natural language (NL) prompting:

\begin{packeditemize}
\item First, there is a \textit{gulf of execution}~\cite{hutchinsDirectManipulationInterfaces1986}: users may not have clear preferences in mind at the outset, and articulating nuanced preferences in language is notoriously difficult~\cite{zamfirescu-pereiraWhyJohnnyCant2023}. Even when users do have a preference, it is still difficult for them to align model output with their expectations for the degree to which that preference is expressed, relying on vague modifiers (``more concise,'' ``not that concise'') to communicate subtle adjustments.
\item Second, there is a \textit{gulf of evaluation}~\cite{normanCognitiveEngineering1986a}: as specified preferences accumulate across turns, the linear chat history grows long and unwieldy to navigate.
Users struggle to compare iterations or attribute differences in the output to particular instructions, making it difficult to decide what to do next~\cite{geroSupportingSensemakingLarge2024,zhangNavigatingFogHow2025}.
\end{packeditemize}
% Our goal is to reduce these two gulfs when users control subjective preferences.\looseness=-1

% \input{figures/fig-example}
\revision{Graphical user interfaces (GUIs)~\cite{hutchinsDirectManipulationInterfaces1986a} have long helped reduce the interaction gulfs by turning users' goals into controllable widgets (\eg dropdowns, toggles, and sliders).
In human-LLM interaction, prior work has used widgets to make prompts more structured and reusable~\cite{macneilPromptMiddlewareMapping2023a,aminPromptCanvasComposablePrompting2025,wangPromptCharmTexttoImageGeneration2024,heContextSteeringControllable2025}. 
However, these systems typically allow users to adjust only predefined dimensions, such as the intensity of a particular attribute using sliders~\cite{boSteerableChatbotsPersonalizing2025}, and provide limited support for inspecting how a given widget influences the generated output.
In this paper, we introduce a more flexible and transparent interaction technique that jointly mitigates these two gulfs in preference control by combining the controllability of GUI widgets with the flexibility of natural-language adjustments (\figref{fig:teaser})}.\looseness=-1

Specifically, we make the following contributions:

\begin{packedenumerate}
    \item We propose \tool (\secref{sec:problem_formulation}), a novel interaction technique that allows users to reify arbitrary preference expressions in their prompts into diverse GUI widgets (\eg toggles, dropdowns, steppers, and sliders). With \tool, users can directly manipulate preferences in their prompts through widgets to control generation (addressing the gulf of execution) and inspect how each preference/widget shapes the output (addressing the gulf of evaluation). 
    \item We build an example system (\secref{sec:system_design}) that instantiates \tool. This system automatically detects and transforms preference expressions in a prompt into GUI widgets, enabling users to configure these widgets to steer generation, inspect their effects on the output via highlights, and track iterations along with widget configurations through a node-link diagram.
    \item We introduce a new LLM decoding algorithm that enables \tool (\secref{sec:technical_framework}). This technical approach modularizes prompts into controllable attributes, enabling independent control over specific preferences to steer model generation and the attribution of generated spans to their corresponding widgets.
    \item We conduct a user study (\secref{sec:user_study}) comparing our method with NL prompting alone on controlled iterative content-creation tasks (\(N=12\)). The results showed that our approach helped users more precisely achieve the target preferences and that they perceived it as more controllable and transparent.
\end{packedenumerate}

In~\secref{sec:discussion}, we discuss how \tool addresses the two gulfs and explore its broader applications beyond the example system presented in this paper.\looseness=-1 
% including the implications of our techniques in the co-creation settings
% JRz: include small sentence to hint at discussion items that handle common critiques (like why is this doing single prompts and not collaboration)
    
% \chao{why do we need to purely rely on models to create artifacts instead of co-writing in a text editor?}

% \chao{My current intro is a bit long.}

% \tool identifies preferences expressed in the prompt, suggests additional relevant preference dimensions, and supports adding new preferences during exploration. These preferences are transformed into interactive controls (\eg dropdowns, toggles, sliders) that users can manipulate directly. To bridge evaluation, \tool supports inspecting the impact of preferences and comparing iterations as paired states of (prompt settings, generated text). Users can switch between versions, compare outputs side-by-side, and relate differences back to the preference settings that produced them.

\section{Background and Related Work}
% We situate \tool within prior work on the challenges of iterative prompting, interfaces that blend language with direct manipulation, and technical methods for steering LLMs.

\subsection{Challenges in Iterative Prompting}
Prompting is expressive: users can describe goals, constraints, and preferences in everyday language, making LLMs easy to access for open-ended writing tasks. 
However, natural language is also an ambiguous interaction medium~\cite{gaoDataToneManagingAmbiguity2015a,fengPromptMagicianInteractivePrompt2024,maAmbigChatInteractiveHierarchical2025,manamTaskLintAutomatedDetection2022}. 
Because many intents and preferences are under-specified, subjective, or context-dependent, users often struggle to convey what they mean in a way the model reliably interprets~\cite{zamfirescu-pereiraWhyJohnnyCant2023}, especially in cases of nuanced preferences like tone and styles. Thus users are often left repeatedly calibrating prompts to approximate an intended target through trial-and-error probing~\cite{mysorePrototypicalHumanAICollaboration2025a}.
% This gulf of execution is especially acute for nuanced preferences---tone, depth, audience fit, or ``how much'' emphasis to place on an aspect—where users may lack stable vocabulary and fall back on vague modifiers (\eg ``more concise,'' ``less formal'') and trial-and-error probing~\cite{mysorePrototypicalHumanAICollaboration2025a}. 
% The result is iterative prompting as calibration: users repeatedly restate, refine, and negotiate constraints to approximate an intended target.
These difficulties are compounded by the turn-by-turn conversational interface common in LLM interaction. 
Iterative prompting scatters preferences across multiple messages, leading to challenges in attribution across drafts~\cite{geroSupportingSensemakingLarge2024,zhangNavigatingFogHow2025,laban_llms_2025}, even with conversation histories~\cite{geroSupportingSensemakingLarge2024} due to lack of reusability.\looseness=-1
% Even LLMs themselves can get lost in multi-turn conversation~\cite{laban_llms_2025}.
% Conversation histories also provide weak support for constraint reuse and version control~\cite{geroSupportingSensemakingLarge2024}: prompts that worked for one draft are hard to extract, generalize, and reapply to create alternative versions. 
% Together, these challenges suggest a tension: while natural language affords broad expressiveness, iterative prompting needs mechanisms that make preferences more stable, inspectable, and adjustable.
% Given the past research, preserving the flexibility of prompting while giving users control over how preferences are applied motivates our work.

% 1. intent is hard to convey/specify in natural language
% 2. tweaking intent is hard (the effect of a specified intent is hard to control and evaluate)
% 3. prompts are hard to reuse across versions
% 4. turn-by-turn conversational interface is bad for version control

% \subsection{Systems to Help Craft Better Prompts}
In order to mitigate these challenges, prior work has designed various tools to aid users during prompt tuning. One line of work automatically suggests effective prompts or directly optimizes user prompts from feedback or examples~\cite{habbaPromptSuiteTaskAgnosticFramework2025,bradePromptifyTexttoImageGeneration2023,liPreferenceGuidedPromptOptimization2026}. Other works build visual environments for prompt development, such as prompt IDEs that support systematic experimentation (\eg organizing variants, running batches, and comparing outputs)~\cite{arawjoChainForgeVisualToolkit2024,kimEvalLMInteractiveEvaluation2024,strobeltInteractiveVisualPrompt2022,mishraPromptAidPromptExploration2025}, as well as visual programming interfaces that help users compose multi-step LLM workflows and inspect intermediate results~\cite{wuAIChainsTransparent2022,zhangVISARHumanAIArgumentative2023,wuPromptChainerChainingLarge2022}. 
% For example, ChainForge supports debugging behaviors at scale by structuring prompt/model combinations and facilitating comparisons across runs~\cite{arawjoChainForgeVisualToolkit2024}. 
% PromptChainer and related systems help users decompose tasks into prompt pipelines, improving reuse and making cascading failures easier to diagnose~\cite{wuPromptChainerChainingLarge2022,wuAIChainsTransparent2022}. 
Overall, these systems strengthen the \textit{engineering} aspect of NL prompting, but they often still treat user preferences as free-form text that must be repeatedly re-specified and interpreted by the model.
Given the past research, preserving the flexibility of NL prompting while giving users control over how preferences are applied motivates our work.
More specifically, we focus on \textit{controllability}: we support the exploration, adjustment, and attribution of subjective dimensions expressed in prompts, making preferences directly manipulable and their effects inspectable for users.
\looseness=-1

\subsection{Blending Language and Direct Manipulation}
\revision{Beyond pure NL prompting, prior work has explored hybrid interfaces that combine natural-language instructions with direct-manipulation or GUI-like controls. 
Many such systems use buttons, tags, or other widgets to execute predefined prompts, package reusable instructions~\cite{gmeinerIntentTaggingExploring2025,richeAIInstrumentsEmbodyingPrompts2025}, or populate prompt templates~\cite{macneilPromptMiddlewareMapping2023a}. 
For example, PromptCanvas~\cite{aminPromptCanvasComposablePrompting2025} enables users to compose and reuse prompt fragments (\eg story elements) to create new prompts for text rewriting. 
DirectGPT compiles users' direct edits to generated text into prompt operations~\cite{masson_directgpt_2024} for text rewriting. 
% Prompt Middleware~\cite{macneilPromptMiddlewareMapping2023a} presents predefined options for the style of writing feedback in a form-based interface; using a slot-based template, these options are combined into a prompt that is then used to generate feedback.
These systems primarily use widgets to organize, reuse, or assemble prompt text before generation. They do not generally support steering of a semantic preference during generation, nor do they reveal how a particular control influences the resulting output.\looseness=-1

A smaller body of work uses widgets as interactive controls for steering generative models.
In image generation, PromptCharm~\cite{wangPromptCharmTexttoImageGeneration2024} supports attention-based manipulation of a database-defined set of modifiers, while AdaptiveSliders~\cite{jainAdaptiveSlidersUseralignedSemantic2025} uses pretrained LoRA modules to provide adjustable visual attributes. 
In text generation, Steerable Chatbots~\cite{boSteerableChatbotsPersonalizing2025} provides a set of predefined opposing dimensions (\eg budget/luxury and kids/adults) by pretraining steering vectors. 
% Textoshop~\cite{massonTextoshopInteractionsInspired2025} allows users to adjust the intensity of a writing style by appending corresponding meta-prompts (\eg ``\textit{Apply the following writing style settings: formality: 2/7}'') to the input.\looseness=-1

Together, these systems demonstrate the value of pairing language with direct manipulation. \tool extends this direction by supporting more flexible preference control and adding an evaluative component that has received limited attention in prior text-generation systems. 
Specifically, rather than restricting interaction to dimensions predefined by developers, databases, or pretrained modules, \tool allows users to turn arbitrary, context-specific preference expressions in their own prompts into a diverse set of manipulable widgets. 
It further provides span-level attributions that reveal how each widget influences specific parts of the generated text. This flexible and inspectable interaction is enabled by a new training-free, decoding-time steering method, rather than LoRA modules or steering vectors that require pretraining.
We summarize these distinctions between \tool and the most closely related systems in Table~\ref{tab}.}

\begin{table}[t]
\centering
\footnotesize
\caption{\formatcaption{Comparison of \tool with closely related systems that combine prompts with direct manipulation.}{\tool (ours) supports training-free steering of arbitrary user-defined preference spans and reveals each widget's influence through span-level attribution.}}
\label{tab}
\begin{tblr}{
    width=\columnwidth,
    colspec={
        Q[l,wd=0.30\columnwidth]
        Q[l,wd=0.11\columnwidth]
        X[l]
        Q[l,wd=0.18\columnwidth]
    },
    colsep=2pt,
    % rowsep=2pt,
    row{1}={font=\bfseries},
}
\toprule
System & Domain & Steering & Attribution \\
\midrule
\textbf{Ours} & Text & Training-free; arbitrary spans & Span-level \\
PromptCanvas~\cite{aminPromptCanvasComposablePrompting2025} 
& Text 
& \xmark{} 
& \xmark{} \\
DirectGPT~\cite{masson_directgpt_2024} 
& Text 
& \xmark{} 
& \xmark{} \\
Steerable Chatbots~\cite{boSteerableChatbotsPersonalizing2025} 
& Text 
& Pretrained steering vectors 
& \xmark{} \\
PromptCharm~\cite{wangPromptCharmTexttoImageGeneration2024} 
& Image 
& Database-defined modifiers 
& Attention \\
AdaptiveSliders~\cite{jainAdaptiveSlidersUseralignedSemantic2025} 
& Image 
& Pretrained LoRA 
& \xmark{} \\
% Textoshop~\cite{massonTextoshopInteractionsInspired2025} 
% & Text 
% & Meta-prompt-based 
% & \xmark{} \\
\bottomrule
\end{tblr}
\end{table}

\subsection{Steering Large Language Models}
\label{sec:steering_rw}
Enabling \tool requires mechanisms that let users steer model generation and inspect outputs based on their preferences through GUI controls at interaction time.
% A key challenge is isolating the influence of an individual preference and enforcing a user-selected setting while other aspects of the prompt remain fixed. 
Prior steering approaches such as parameter-efficient fine-tuning (\eg LoRA)~\cite{huLoRALowRankAdaptation2021} and activation steering~\cite{boSteerableChatbotsPersonalizing2025} can offer strong control, but typically require additional training, curated data, or offline preparation. This limits their applicability in \revision{controlling arbitrary preference expressions}. 
In HCI systems, the most common interaction-time strategy is instead to encode values or intensities using meta-instructions within a prompt (\eg ``on a scale from 0 to 10'' in Textoshop~\cite{massonTextoshopInteractionsInspired2025}).
However, prompt-based scales can be inconsistently interpreted and may not align with human perception~\cite{zhangPreciseAttributeIntensity2025,labrooFunnyPersuasiveNot2026,haldarRatingRouletteSelfInconsistency2025,zhouBatchCalibrationRethinking2024}. 
We instead introduce a new LLM decoding algorithm that isolates and controls each attribute by directly modulating the model's next-token probability distribution during inference.
% , motivated by the observation that adding an attribute phrase changes the relative likelihood of candidate next tokens~\cite{heContextSteeringControllable2025}.
This method supports (1) real-time adjustment of multiple independent attributes via GUI controls and (2) attribution by linking changes in token probabilities to the corresponding controls. 
We detail our technical method \revision{and compare it with prompt-based scaling} in~\secref{sec:technical_framework}.\looseness=-1

\begin{table*}
\footnotesize
\centering
\caption{\formatcaption{Taxonomy of the preference space.}{The taxonomy organizes the preference space based on common dimensions identified in prior literature~\cite{leeAligningThousandsPreferences2024,malaviyaContextualizedEvaluationsJudging2025,kimCUPIDEvaluatingPersonalized2025,zhou_beyond_2025,zhu_show_2025}, along with the typical attribute types through which each dimension manifests.}}
\label{tab:preference_taxonomy}
\begin{tblr}{
  colspec = {X[1.5,l] X[2.2,l] X[2.2,l] X[1.5,l] X[1.2,l]},
  hline{1,Z} = {0.8pt, solid},
  row{1} = {font=\bfseries},
  % hlines
}
Preference Dimension & Description & Examples & Typical Attribute(s) & Sources \\
\hline
\textbf{Format \& Structure} & 
How the output is visually, logically, or syntactically organized. & 
``as a poem,'' ``in JSON,'' ``use bullet points,'' ``two columns'' & 
Categorical, Binary &
\cite{leeAligningThousandsPreferences2024,malaviyaContextualizedEvaluationsJudging2025,kimCUPIDEvaluatingPersonalized2025,zhu_show_2025,zhou_beyond_2025} \\
\textbf{Tone \& Style} & 
The emotional resonance, voice, or stylistic flavor of the writing. & 
``funny,'' ``professional,'' ``academic,'' ``enthusiastic'' & 
Categorical, Continuous &
\cite{leeAligningThousandsPreferences2024,malaviyaContextualizedEvaluationsJudging2025,kimCUPIDEvaluatingPersonalized2025} \\
\textbf{Audience \& Persona} & 
Who the text is tailored for, or the role the LLM should assume. & 
``explain to a 5-year-old,'' ``for domain experts,'' ``act as a lawyer'' & 
Categorical, Binary &
\cite{leeAligningThousandsPreferences2024,malaviyaContextualizedEvaluationsJudging2025,zhu_show_2025,zhou_beyond_2025} \\
\textbf{Length \& Conciseness} & 
Constraints on the physical size or detail level of the output. & 
``short,'' ``under 200 words,'' ``3 paragraphs,'' ``concise'' & 
Numeric, Continuous &
\cite{leeAligningThousandsPreferences2024,malaviyaContextualizedEvaluationsJudging2025,zhou_beyond_2025} \\
\textbf{Content Constraints} & 
Specific rules about what information to include or exclude. & 
``no jargon,'' ``include references,'' ``focus on the impact'' & 
Categorical, Binary &
\cite{malaviyaContextualizedEvaluationsJudging2025,zhou_beyond_2025} \\
\end{tblr}
\Description{Full-width table titled ``Taxonomy of the preference space.'' The table has five columns: Preference Dimension, Description, Examples, Typical Attribute(s), and Sources. It organizes user preferences for LLM outputs into five dimensions. Format and Structure refers to how output is visually, logically, or syntactically organized, with examples such as ``as a poem,'' ``in JSON,'' ``use bullet points,'' and ``two columns,'' and is associated with categorical and binary attributes. Tone and Style refers to emotional resonance, voice, or stylistic flavor, with examples such as ``funny,'' ``professional,'' ``academic,'' and ``enthusiastic,'' and is associated with categorical and continuous attributes. Audience and Persona refers to who the text is tailored for or what role the model should assume, with examples such as ``explain to a 5-year-old,'' ``for domain experts,'' and ``act as a lawyer,'' and is associated with categorical and binary attributes. Length and Conciseness refers to constraints on output size or detail level, with examples such as ``short,'' ``under 200 words,'' ``3 paragraphs,'' and ``concise,'' and is associated with numeric and continuous attributes. Content Constraints refers to rules about what information to include or exclude, with examples such as ``no jargon,'' ``include references,'' and ``focus on the impact,'' and is associated with categorical and binary attributes. The final column cites prior literature supporting each dimension.}
\end{table*}

\section{What is Malleable Prompting?}
\label{sec:problem_formulation}

Here, we formalize the concept of \tool, which serves to guide the interaction design described in \secref{sec:system_design} and the technical method introduced in \secref{sec:technical_framework}.
% \yiren{todo: revisit the designs and make sure the rationales between design decisions are strongly justifiable}

\subsection{Preferences and Attributes}
A prompt can be composed of two distinct parts: a \emph{task specification} (the core objective) and \emph{preference expressions} (constraints on execution)~\cite{zhu_show_2025}.
% While task specifications are usually explicit, preferences regarding style, format, or tone are often embedded implicitly.
For example, in the prompt ``\textit{Write a friendly Slack announcement to team members about a new company-wide social event in a brief paragraph},'' the task is the explanation itself, while ``\textit{team members}'' (audience), ``\textit{brief}'' (length), ``\textit{friendly}'' (tone), and ``\textit{Slack announcement}'' (format) act as constraints that shape \textit{how} that task is executed.
Here, we define an \emph{attribute} as a specific preference constraint embedded in the prompt that can be parameterized. A \emph{control} is the GUI widget that exposes this attribute for manipulation.
The core problem \tool addresses is transforming a static prompt into a set of independently manipulable attributes, exposed as controls that users can adjust interactively.
\looseness=-1

To systematically model how users constrain LLM outputs via preference expressions, we conceptualize a \emph{preference space} based on common dimensions identified in literature~\cite{leeAligningThousandsPreferences2024,malaviyaContextualizedEvaluationsJudging2025,kimCUPIDEvaluatingPersonalized2025,zhou_beyond_2025,zhu_show_2025}. We categorize these into five primary dimensions (Table \ref{tab:preference_taxonomy}): Format \& Structure (structural organization), Tone \& Style (stylistic flavor), Audience \& Persona (target reader or assumed persona), Length \& Conciseness (conciseness constraints), and Content Constraints (specific inclusions/exclusions).
We use this taxonomy to inform the design and guide the implementation of the system (see \secref{sec:system_design}). While not exhaustive, this taxonomy captures the most common preference dimensions that users actively control. 
Importantly, our system remains flexible enough to accommodate new, emergent preference dimensions beyond this foundational set.

To incorporate these high-level preference dimensions in an interactive technique, we must map them to specific, manipulable data structures.
Drawing from established taxonomies in measurement theory~\cite{stevens_theory_1946}, we classify attributes into four types based on their value properties: categorical, numeric, binary, and continuous.\looseness=-1

\begin{packeditemize}
\item \textbf{Categorical:} A choice among mutually exclusive nominal options (\eg output format: ``HTML,'' ``Markdown,'' ``Plain Text'').
\item \textbf{Numeric:} A countable, discrete quantity (\eg ``3 paragraphs'').
\item \textbf{Binary:} A presence/absence constraint (\eg ``include a title'' vs.\ none; ``use bullet points'' vs.\ none).
\item \textbf{Continuous:} An ordinal intensity along a stylistic spectrum (\eg ``more formal,'' ``less concise'').
\end{packeditemize}

Table \ref{tab:preference_taxonomy} illustrates the typical attributes through which each preference dimension manifests.

\subsection{Problem Formulation}
\label{sec:math_formulation}

Formally, let $P_0$ represent the initial task specification, $\mathcal{A}$ represent the set of attributes for preference expressions, partitioned into four subsets: $\mathcal{A}^{\text{cat}}$, $\mathcal{A}^{\text{num}}$, $\mathcal{A}^{\text{bin}}$, and $\mathcal{A}^{\text{cont}}$. 
We model the full prompt $P$ as the composition of $P_0$ with these attribute sets:
\begin{equation}
P_{\mathcal{A}} = P_0 \oplus (\mathcal{A}^{\text{cat}} \cup \mathcal{A}^{\text{num}} \cup \mathcal{A}^{\text{bin}} \cup \mathcal{A}^{\text{cont}})
\end{equation}
Here, $\oplus$ denotes a composition operator that integrates attributes into a parameterized prompt.
Each attribute corresponds to a phrase span in the prompt.
Discrete attributes (\ie categorical, numeric, binary) take on distinct values, while continuous attributes represent intensity along a semantic scale (\eg ``concise'').
Given a parameterized prompt $P_{\mathcal{A}}$, generation produces an output $Y \sim p(\cdot \mid P_{\mathcal{A}})$.
\tool aims to support interactive exploration over $\mathcal{A}$ by enabling users to vary attributes and observe the corresponding changes in $Y$.\looseness=-1

\subsection{Design Goals}
\label{sec:design_goals}
To bridge the gulfs of execution and evaluation identified in \secref{sec:intro}, and to operationalize the formal definition of \tool into an interactive system, we derive two core design goals (DG):

\namedparagraph{DG1 (Execution): Reify Preferences as GUI Widgets to Steer Generation}
To realize the model of $P_{\mathcal{A}}$, the system should be able to help users identify controllable preferences embedded in a user's prompt ($P$) and reify them into appropriate GUI widgets (elements of $\mathcal{A}$). 
Moreover, because users often do not know their full set of preferences a priori and preferences emerge in response to seeing outputs~\cite{zamfirescu-pereiraWhyJohnnyCant2023}, the system should allow users to add new constraints that expand $\mathcal{A}$ over time and can suggest relevant unstated preferences.\looseness=-1

\namedparagraph{DG2 (Evaluation): Reveal Widget Effects on Output to Support Attribution}
In standard chat interfaces, the link between a prompt tweak and the resulting output change is often opaque. To support the evaluation of outputs ($Y$) so users can make informed iterations, the system must explicitly map the relationship between control adjustments and text variations. It should help users track how specific attribute values influence specific text spans and organize drafts around configurations (states of $\mathcal{A}$) rather than linear conversation history, enabling users to easily compare, revert, and branch iterations.

\section{System Design}
\label{sec:system_design}
\tool is an interactive prompting technique that allows users to reify preference expressions in their prompts into embedded GUI controls.
With \tool, users can directly manipulate these preferences through widgets to steer generation (\textbf{DG1}) and inspect how each preference shapes the output (\textbf{DG2}).
To illustrate this, we instantiated \tool within a research prototype guided by the two design goals and established principles of direct manipulation~\cite{hutchinsDirectManipulationInterfaces1986,masson_directgpt_2024,shneiderman_direct_1981}.
We discuss broader applications of \tool across several other interactive systems in \secref{sec:broader_applications}.
In this section, we detail the prototype's core functionalities through a user scenario.\looseness=-1

% \chao{we still followed principle in direct manipulation.}

\begin{figure}
  \centering
  \includegraphics[width=\linewidth]{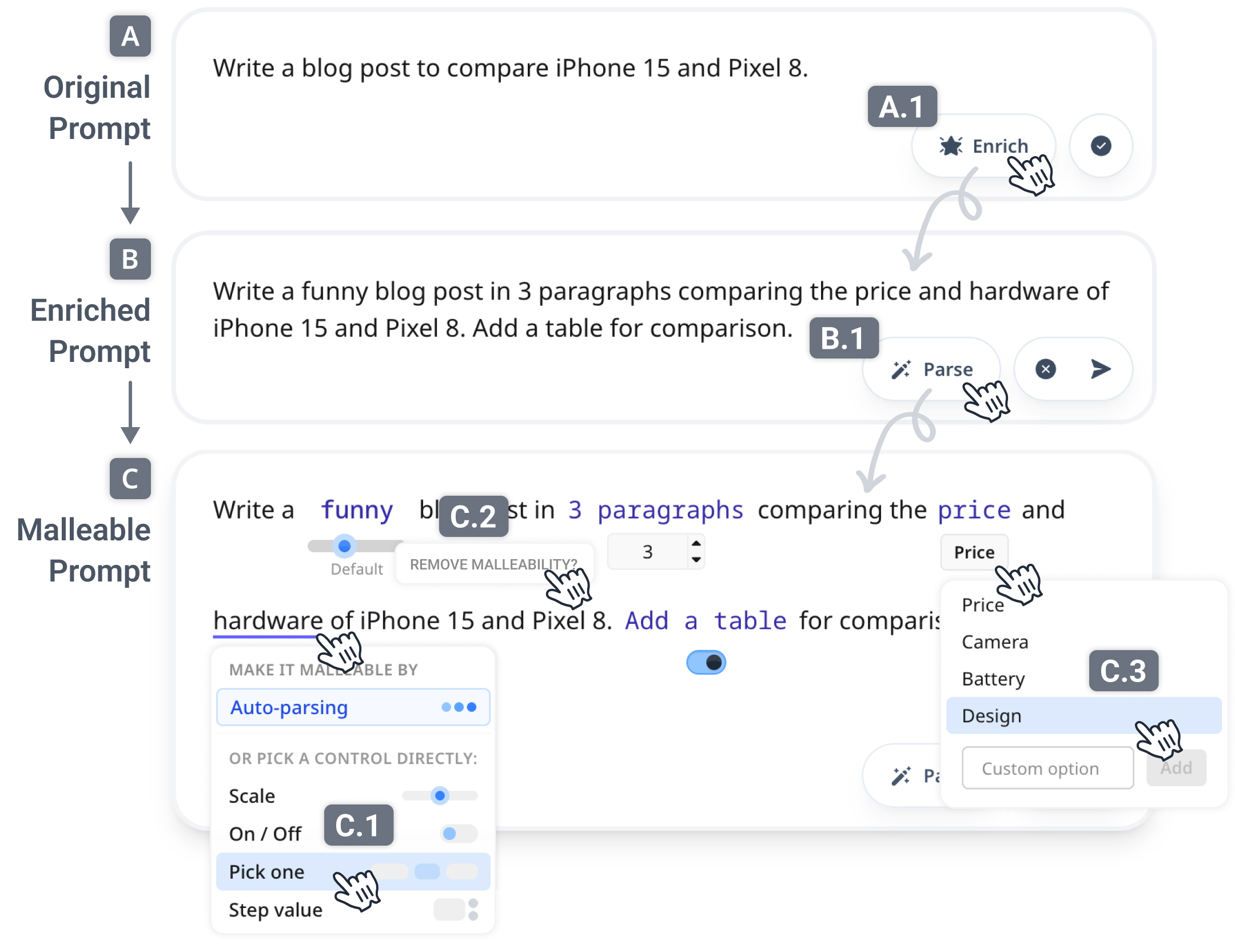}
  \caption{\formatcaption{Making prompts malleable.}{\revision{Users can enrich an initial prompt with suggested preferences (A.1), automatically parse controllable attributes and convert them into widgets or manually bind selected text to a widget (C.1). Users can remove widgets via ``Remove Malleability'' (C.2). For categorical attributes, the system also suggests alternative values for later iterations (C.3).}}}
  \Description{Three stacked interface panels illustrate a workflow for transforming a plain prompt into an interactive, malleable prompt with GUI controls. In the top panel, the prompt reads, ``Write a blog post to compare iPhone 15 and Pixel 8,'' with a button labeled ``Enrich'' on the right; label A marks this stage. A curved arrow points to the middle panel, where the prompt has been automatically expanded to ``Write a funny blog post in 3 paragraphs comparing the price and hardware of iPhone 15 and Pixel 8. Add a table for comparison.'' A button labeled ``Parse'' appears on the right, and label B marks this stage. Another arrow points to the bottom panel, where parts of the prompt have been converted into editable widgets: ``funny'' is paired with a slider, ``3 paragraphs'' with a numeric stepper, ``price'' with a pick-one dropdown, ``Add a table'' with an on/off toggle, and ``hardware'' is selected for control assignment. A pop-up menu labeled ``Make it malleable by'' shows options such as ``Auto-parsing,'' ``Scale,'' ``On / Off,'' ``Pick one,'' and ``Step value''; label C highlights this control-selection menu. A second dropdown for the ``price'' term lists alternatives including ``Price,'' ``Camera,'' ``Battery,'' and ``Design,'' with ``Design'' highlighted; label D marks this pick-one menu. Hand cursors and arrows emphasize the sequential interaction flow from enrichment, to parsing, to direct widget assignment.}
  \label{fig:prompt}
  \vspace{-10pt}
\end{figure}

\subsection{User Scenario}
Assume a user, Alice, wants to draft and iteratively refine a blog post using an LLM.
To begin, she enters the prompt: ``\textit{Write a blog post in 3 paragraphs comparing iPhone~15 and Pixel~8.}'' into the system.

\namedparagraph{Enriching Prompts with Suggested Preferences}
Initially, Alice is unsure of the specific style and constraints for her post.
She clicks the star {\small\faStar} icon (\figref{fig:prompt}A.1), and the system automatically enriches her initial prompt with several preference expressions to help her get started (\textbf{DG1}).
The prompt becomes: ``\textit{Write a funny blog post in 3 paragraphs comparing the price and hardware of iPhone~15 and Pixel~8. Add a table for comparison.}''
Alice decides to proceed with this enriched version.\looseness=-1

\namedparagraph{Parsing Prompts into Controllable Widgets}
Next, Alice clicks the auto-parsing {\small\faMagic} icon (\figref{fig:prompt}B.1). The system automatically extracts controllable attributes from the text and reifies them into GUI controls (\textbf{DG1}). It converts ``\textit{funny}'' into a continuous tone slider, ``\textit{3 paragraphs}'' into a numeric length stepper, ``\textit{price and hardware}'' into a categorical dropdown, and ``\textit{Add a table for comparison}'' into a binary formatting toggle.
Alice also notices that she can manually highlight any text span to bind it to a specific widget type: slider, toggle, dropdown, or stepper (\textbf{DG1}). 
She replaces the auto-generated dropdown for ``\textit{price and hardware}'' with two distinct categorical dropdowns: one for ``\textit{price}'' and another for ``\textit{hardware}'' (\figref{fig:prompt}C.1).
For these controls, the system suggests alternative values (\eg \textit{aesthetics, camera, battery}) to populate the dropdown menus, which Alice can easily include or discard.\looseness=-1

\namedparagraph{Steering Generation via Widget Manipulation}
With the controls configured, Alice explores different preference states (\textbf{DG1}).
She shifts the tone slider toward a higher intensity of ``\textit{funny},'' leaves the table toggle enabled, changes ``\textit{hardware}'' to the more specific ``\textit{camera}'' option, and changes the comparison dimension from ``\textit{price}'' to ``\textit{design}'' via the dropdowns (\figref{fig:prompt}C.2).
Upon clicking the generate {\small\faPaperPlane} icon, the system leverages this specific attribute configuration to steer the LLM, producing a draft that adopts a humorous tone and formats the design and hardware comparison as a table.
\looseness=-1

\namedparagraph{Inspecting Widget Influence on the Output}
To understand exactly how her configuration shaped the draft, Alice clicks ``\textit{funny}'' to inspect its specific effects (\textbf{DG2}). The system responds by highlighting the impacted sentences in the output. Within those sentences, specific words and phrases encouraged by ``\textit{funny}'' (\eg ``\textit{Oh boy,}'' ``\textit{national anthem,}'' ``\textit{a spoon made of glitter}'') are highlighted with a green background, while discouraged formal or dry terms are marked with a red background. She similarly inspects the ``\textit{camera}'' dropdown and the formatting toggle, quickly verifying how those specific constraints localized their impact within the text. By exploring these granular highlights across different widgets, Alice can make informed decisions about whether her current configuration aligns with her intent for the next iteration.\looseness=-1

\begin{figure}
  \centering
  \includegraphics[width=\linewidth]{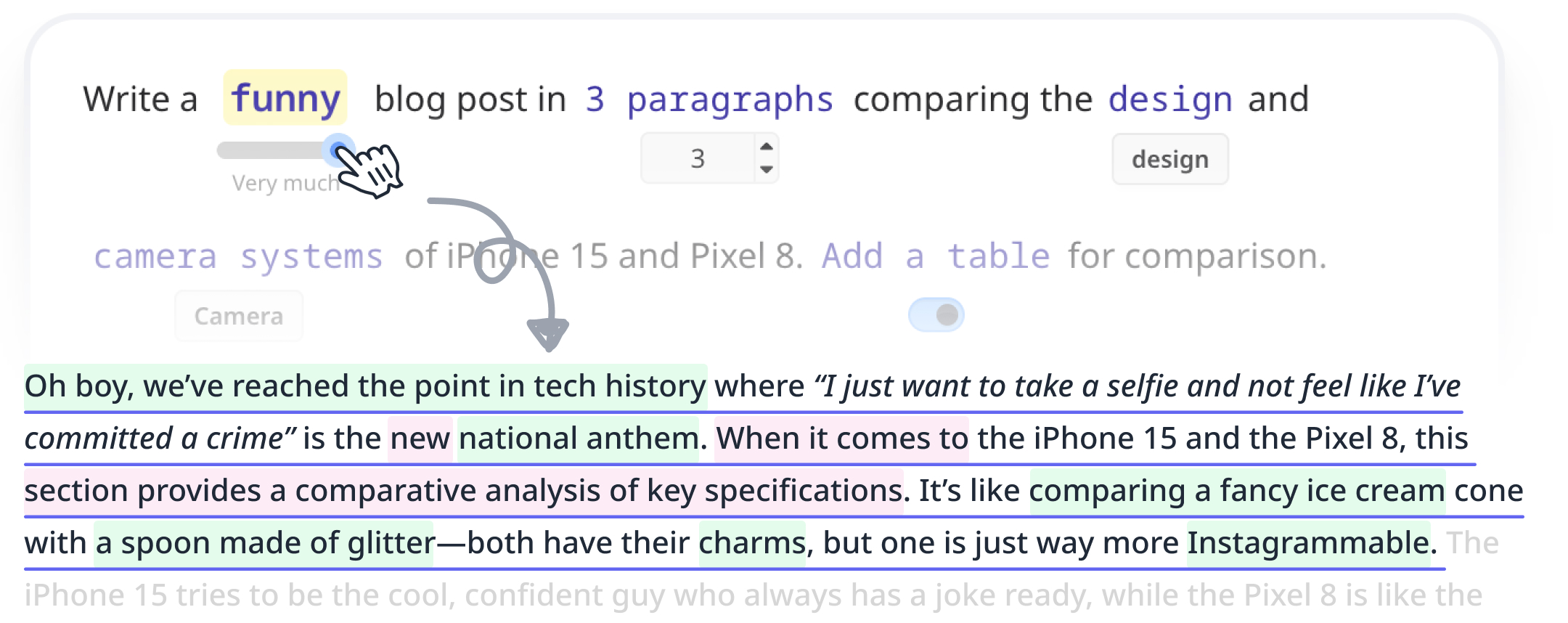}
  \caption{\formatcaption{Inspecting widget effects on generated text.}{Clicking a widget reveals its localized influence on the output: affected sentences are underlined, positively induced phrases are highlighted in green, and discouraged wording is highlighted in red.}}
  \Description{A text-generation interface showing a prompt that reads, ``Write a funny blog post in 3 paragraphs comparing the design and camera systems of iPhone 15 and Pixel 8. Add a table for comparison.'' The words ``funny,'' ``3 paragraphs,'' ``design,'' ``camera systems,'' and ``table'' are highlighted in purple, with controls such as a funny slider set high, a paragraph count selector set to 3, and a design tag. A hand cursor points to the humor slider, and a doodled arrow and circle connect the prompt to the generated paragraph below. The output begins with a humorous comparison of the two phones, and several phrases are highlighted and underlined to show how specific prompt settings influenced the generated text.}
  \label{fig:inspect}
  \vspace{-10pt}
\end{figure}

\namedparagraph{Iterating Content via Widget Configuration}
Based on these highlights, Alice refines her content by configuring her malleable prompts.
She types a new preference expression ``\textit{sarcastic}'' into the text and converts it into a new continuous slider control (\figref{fig:teaser}) to pivot the blog post's overall flavor (\textbf{DG1}).
Accordingly, she decreases the intensity of the ``\textit{funny}'' tone slider to better balance the two comedic styles.
Over the next few minutes, Alice configures her malleable prompt multiple times, tweaking slider intensities, adjusting the number of paragraphs, toggling the comparison table, and swapping dropdown values to generate and evaluate several alternative versions of the blog post.
\looseness=-1

\begin{figure}
  \centering
  \includegraphics[width=\linewidth]{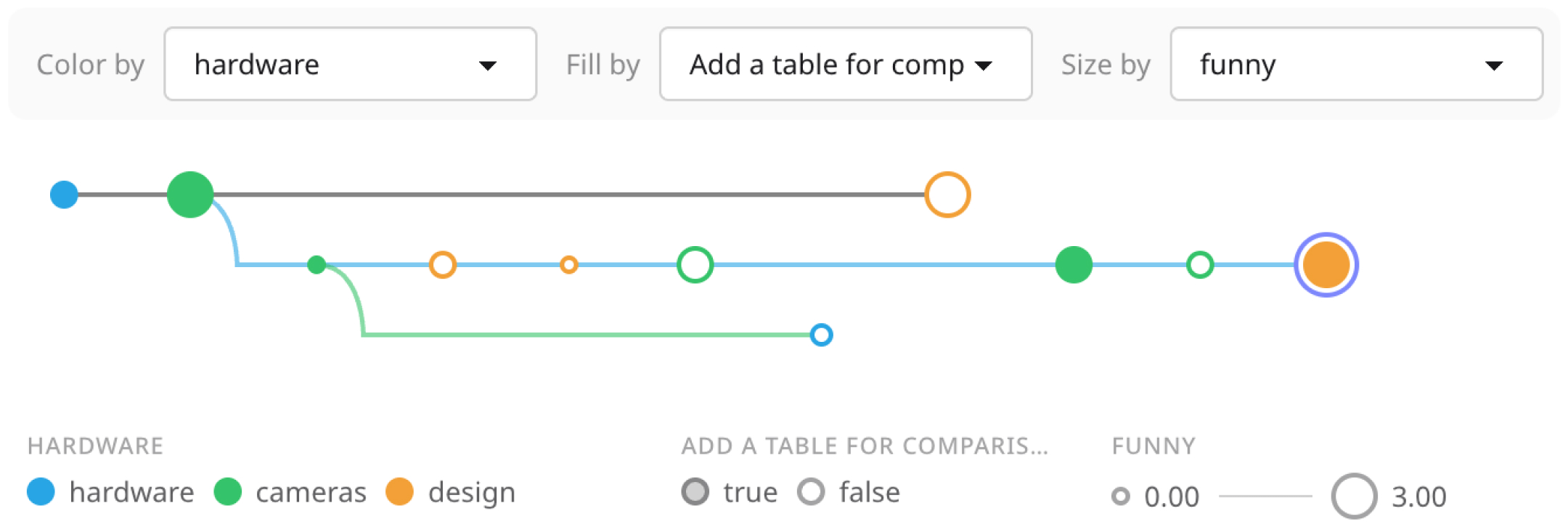}
  \caption{\formatcaption{Tracking prompt iterations with a node-link graph.}{Versions generated with the same prompts and widgets appear along the same horizontal branch. Adding new preferences or widgets creates a new vertical branch. Visual encodings map widget values to node color for categorical attributes, node fill for binary attributes, and node size for continuous or numerical attributes.}}
  \Description{A visualization interface showing how prompt attributes of each iterations are encoded visually. Three dropdown menus at the top read ``Color by hardware,'' ``Fill by Add a table for comp...,'' and ``Size by funny.'' Below them, a node-link style chart displays colored circles connected by horizontal and stepped lines. Circle color represents hardware-related categories, with blue for hardware, green for cameras, and orange for design. Circle fill indicates whether the attribute ``Add a table for comparison'' is true or false, using filled versus hollow markers. Circle size represents the degree of ``funny'' on a scale from 0.00 to 3.00, with larger circles indicating higher values. Legends at the bottom explain the color, fill, and size mappings.}
  \label{fig:graph}
  \vspace{-10pt}
\end{figure}

\namedparagraph{Managing Iterations via the Version Graph}
After several rounds of iterations, Alice uses the version graph visualization to make sense of her iteration history (\textbf{DG2}). 
The graph displays her generated alternatives as a node-link diagram, arranged chronologically from left to right (\figref{fig:graph}). 
Nodes representing versions generated with the same prompts and widgets, though possibly with different values, appear on the same horizontal branch. 
Conversely, adding new preferences and widgets (like the ``\textit{sarcastic}'' slider) creates a new vertical branch. 
Different attributes can be mapped to different visual encoding channels. For example, Alice maps the categorical ``\textit{hardware}'' dimension to node color, the binary ``\textit{table}'' constraint to node fill (solid vs.\ hollow), and the continuous ``\textit{funny}'' tone to node size. 
She can click any node to revert the prompt to that exact state. By comparing prompt configurations and their corresponding outputs, Alice identifies her preferred version and exports it.\looseness=-1

% \input{figures/fig-revise}

% \namedparagraph{Fine-tuning Text via Local Steering}
% Now reviewing her chosen draft, Alice clicks ``\textit{camera}'' to inspect its footprint in the text (\textbf{DG2}). 
% Guided by the resulting indigo highlights, she notices that the camera comparison section remains a bit dry. 
% To address this locally, she highlights that specific text span and inputs a revision prompt: ``\textit{make it punchier}'' (\figref{fig:revise}).
% The system reifies this local instruction into an intensity slider (\textbf{DG1}).
% As Alice drags the slider, the selected text updates in real time, allowing her to quickly compare varying degrees of ``punchiness'' without altering the rest of the document. 
% Satisfied with this final polish, she exports her finished post.

\subsection{Implementation Notes}
The frontend of our system is built using Next.js.
The core technical method (described in \secref{sec:technical_framework}), responsible for modularizing prompts into controllable attributes for manipulation and attribution, and steering a locally deployed language model's generation during decoding, is implemented in Python. 
Other text processing components, including enriching prompts, parsing attributes, and generating options, are powered by prompting the \texttt{GPT-5.3} model via the OpenAI API.
Prompts used are provided in Appendix~\ref{appendix:technical_details}.\looseness=-1

\section{Technical Methods}
\label{sec:technical_framework}
In this section, we present the technical method underlying \tool for addressing the problem formulated in \secref{sec:math_formulation}.
The interaction in \tool requires the ability to steer model generation according to customized widgets and their values, for arbitrary preference expressions selected from the prompt.
This requirement makes many steering approaches used in prior HCI systems infeasible, such as parameter-efficient fine-tuning methods~\cite{huLoRALowRankAdaptation2021,jainAdaptiveSlidersUseralignedSemantic2025} and activation steering~\cite{boSteerableChatbotsPersonalizing2025}, because they typically rely on additional training, curated data, or offline preparation.
Instead, our method builds on the observation that adding or removing a phrase changes token probability distribution during inference~\cite{heContextSteeringControllable2025}, and that these changes can be leveraged to steer generation at decoding time.\looseness=-1

Based on this insight, we introduce a training-free method with two key components: (1) steering generation based on user-adjusted configurations for each attribute, and (2) attributing the effect of each attribute to corresponding textual spans in the output.
\looseness=-1

% \ly{$P$, $P_0$ and $A$ are not defined here}

\subsection{Prompt Modularization}
\label{sec:prompt_steering}
In this component, we first instruct \texttt{GPT-5.3} to extract controllable attributes $\mathcal{A}$ in a given prompt $P$ and the corresponding phrase span (preference expressions) associated with each attribute.
We then split $P$ into an initial task specification $P_0$ (\eg ``\textit{write an email ...}'') and a set of attributes $\mathcal{A}$ (\eg ``\textit{in a formal tone}'' and ``\textit{using bullet points}'') for preference expressions.

For each attribute $a \in \mathcal{A}$, we isolate its specific contribution to the generation. At each decoding step $i$, we estimate the token-level influence of $a$ as the difference in log-probabilities between prompting with and without the attribute:
\begin{equation}
F_{a,P_0}(x_i) = \log p(x_i \mid x_{<i}, a, P_0) - \log p(x_i \mid x_{<i}, \varnothing, P_0).
\end{equation}
The value $F_{a,P_0}(x_i)$ quantifies the degree to which including $a$ shifts the likelihood of token $x_i$ in the current context.

For binary and continuous attributes, we expose a scalar control $\lambda_a$ that scales the attribute's intensity during decoding:\looseness=-1
\begin{equation}
\log p_{\lambda_a}(x_i \mid x_{<i}, a, P_0) = \log p(x_i \mid x_{<i}, \varnothing, P_0) + \lambda_a F_{a,P_0}(x_i).
\end{equation}
By default, $\lambda_a=1$. Setting $\lambda_a=0$ recovers the original distribution without the attribute, while varying $\lambda_a$ amplifies or attenuates its effect. We restrict $\lambda_a \in \{1, 0\}$ for binary attributes to emulate an on/off toggle. For continuous attributes, we map $\lambda_a \in [0, 3]$ to a 7-point slider with increments of 0.5.
\revision{We selected this range based on the technical evaluation described below.}
Appendix~\ref{appendix:examples} presents example outputs generated with different values of $\lambda$ for continuous attributes.
Conversely, categorical and numeric attributes represent discrete constraints that require strict adherence, such as a specific language or word count. We handle these by enforcing the user's choice through direct string substitution of $a$ in $P$.\looseness=-1

To account for the combined effect of all attributes, we define the final modulated distribution. Let $\mathcal{A}_{\text{sub}} = \mathcal{A}^{\text{cat}} \cup \mathcal{A}^{\text{num}}$ be the set of attributes handled via substitution, and $\mathcal{A}_{\text{mod}} = \mathcal{A}^{\text{bin}} \cup \mathcal{A}^{\text{cont}}$ be those handled via modulation. We combine their effects as follows:
\begin{equation}
\log p_{\boldsymbol{\lambda}}(x_i \mid x_{<i}, \mathcal{A}) = \log p(x_i \mid x_{<i}, P_{\text{sub}}) + \sum_{a \in \mathcal{A}_{\text{mod}}} \lambda_a F_{a,P_0}(x_i),
\end{equation}
where $P_{\text{sub}}$ represents the prompt formed by substituting all attributes of $\mathcal{A}_{\text{sub}}$ in $P_0$ with their user-selected values.

% To account for the joint effect of multiple attributes, we combine their token-level influences at decoding step $i$:
% \begin{align}
% \log p\left(x_i \mid x_{<i}, \mathcal{A}, P_0\right)
% = &\log p\left(x_i \mid x_{<i}, \mathcal{A}^{\text{cat}} \cup \mathcal{A}^{\text{num}}, P_0\right) \\&+ \sum_{a \in \{\mathcal{A}^{\text{bin}} \cup \mathcal{A}^{\text{cont}}\} } \lambda_a  F_{a,P_0}(x_i).
% \end{align}

This method allows users to steer model generation with their chosen preferences at decoding time without offline preparation or additional training.
We implement it using an open-source model (\texttt{Qwen3-8B}) to enable access to logits and custom decoding.
% with batched inference and KV-cache reuse to mitigate the extra latency. Detailed computation overhead analysis is in Appendix \ref{appendix:computation_overhead}.

\revision{
\namedparagraph{Technical Evaluation}
We evaluated whether our method provides (1) stronger steering and (2) more calibrated control than prompt-based scaling using an LLM-as-a-judge protocol with human validation.
For the baseline, we prepended explicit numeric instructions for each active attribute to control intensity (\eg ``\textit{Apply the following writing style settings: formality: 2/7},'' following Masson et al.~\cite{massonTextoshopInteractionsInspired2025}) and generated outputs using standard decoding.
For our method, we explored $\lambda \in [-10, 10]$.
Using 100 prompts sampled from the \texttt{WildChat-WC\textsubscript{Wr}} dataset~\cite{mysorePrototypicalHumanAICollaboration2025a} and 30 preference combinations created based on the corpus of Lee et al.~\cite{leeAligningThousandsPreferences2024}, we varied the baseline's numeric intensity settings from 1 to 7 and mapped $\lambda$ to the same 1--7 scale to generate outputs. 
We then used an LLM judge (\texttt{GPT-5-mini}) to rate the perceived intensity of each attribute in each output on a 1--7 scale.
We validated the judge on 126 generated outputs independently rated by three researchers, obtaining moderate agreement with human annotations ($\kappa = 0.410$, $p<.001$).\looseness=-1

For each method and dimension, we fit a least-squares regression of the judge score on the intensity level, where the regression coefficient $\beta$ estimates the expected change in perceived attribute intensity for each one-unit increase.
Overall, our steering method achieved greater steering magnitude than the prompt-only baseline across the tested attributes (mean $\beta = 0.61$ vs.\ $0.32$, $p < .001^{***}$).
Moreover, our method produced slopes significantly closer to 1 than the prompt-only baseline (mean $|\beta - 1| = 0.40$ vs.\ $0.69$; paired $t = -5.59$, $p < .001^{***}$), indicating that it more closely approximated the intended intensity scale.
Examining the full $\lambda$ sweep further revealed that steering strength increased substantially until approximately $\lambda = 3.33$, after which performance saturated.
Also considering negative $\lambda$ values were difficult to interpret as user-facing control settings, we therefore map the slider in the final deployed system to a $\lambda$ range of $[0, 3]$ by rounding the empirically identified saturation point of $3.33$ down to $3$.
We provide additional details of the protocol and results in Appendix~\ref{appendix:cos_eval}.\looseness=-1
}

\subsection{Prompt Attribution}
The second component traces which portions of the generated output are influenced by each controllable attribute. We adopt a post-hoc attribution method that provides users with transparency, allowing them to understand the specific impact of their adjustments.\looseness=-1

For discrete attributes handled via substitution in $\mathcal{A}_{\text{sub}}$, we utilize \texttt{GPT-5.3} to identify exact, verbatim substrings within the generated output that correspond to the substituted values in $P_{\text{sub}}$. This lexical matching provides a direct link between the user's choice and the resulting text.
In contrast, attributing influence to continuous attributes in $\mathcal{A}_{\text{mod}}$ is more complex because their effects often interact during decoding; consequently, isolating token-level influence to a single attribute can be complex. Following prior work on model explanation~\cite{chenAlgorithmsEstimateShapley2023,sundararajan2020many}, we thus compute the Shapley attribution that measures the \emph{marginal} effect of an attribute $a$  while accounting for the presence of others.

For each generated token $x_i$, we define the attribution $\phi_a(x_i)$ of attribute $a \in \mathcal{A}_{\text{mod}}$ as the ratio of its probability under the full control vector $\boldsymbol{\lambda}$ versus a counterfactual that removes only that specific attribute:
\begin{equation}
\phi_a(x_i) := \frac{p_{\boldsymbol{\lambda}}(x_i \mid x_{<i}, \mathcal{A})}{p_{\boldsymbol{\lambda}_{\setminus a}}(x_i \mid x_{<i}, \mathcal{A})}.
\end{equation}
In this context, $\boldsymbol{\lambda}_{\setminus a}$ denotes the control vector where the weight for attribute $a$ is set to zero ($\lambda_a = 0$), while all other weights remain at their user-specified levels.

A value of $\phi_a(x_i) > 1$ indicates that attribute $a$ increases the likelihood of token $x_i$ given the context; conversely, $\phi_a(x_i) < 1$ indicates suppression. We visualize contiguous spans of tokens with consistent positive or negative values to help users anticipate how further adjustments to $\lambda_a$ will reshape the output. 
This approach implements the core Shapley principle of marginal contribution while avoiding exponential subset enumeration, which ensures the computation remains feasible during real-time decoding (Appendix~\ref{appendix:computation_overhead}).\looseness=-1

\revision{
\namedparagraph{Technical Evaluation}
We evaluated whether attribution highlights faithfully reflect a slider's effect by testing whether a token's displayed label predicts the direction of its probability change when the corresponding slider is adjusted.
Using the same model and dataset as in \secref{sec:prompt_steering}, we treated intensity level 5 as the reference condition and measured token-level probability changes when varying the $\lambda$ value of an individual attribute to produce higher intensity (7) or lower intensity (3 and 1).

The results showed that the tokens highlighted by our method matched the observed direction of change 81.4\% of the time, compared with 67.2\% for neutral tokens, indicating that the highlights carry meaningful predictive signal. This effect was also attribute-specific: agreement was 81.0\% when adjusting the intensity of a token's corresponding attribute, but only 60.8\% when adjusting a different attribute, near the neutral baseline. 
Together, these results suggest that, although our highlights are post-hoc indicators rather than causal explanations, they provide a useful widget-specific signal of how manipulating a widget changes the output distribution.
We report the full protocol and results in Appendix~\ref{appendix:attribution_eval} and Table \ref{tab:attribution}.}

\section{User Study}
\label{sec:user_study}

We conducted a within-subjects study with 12 participants, comparing \tool with NL prompting.
The study aimed to address the following research questions:

\begin{enumerate}
    \item[\textbf{RQ1}] Does \tool help users better control LLM generation according to their preferences?
    \item[\textbf{RQ2}] Does \tool help users better inspect the effects of preference changes on the output?
    \item[\textbf{RQ3}] How do users appropriate \tool, and what strategies and challenges emerge?
\end{enumerate}

\subsection{Participants}
We recruited 12 participants (7 female, 5 male) aged 24--37 ($M = 26.67$, $SD = 3.68$) via social networks and word of mouth. 
All participants reported using LLM tools (\eg ChatGPT, Claude, or Gemini) for writing and editing tasks on a daily basis. 
% Regarding their typical weekly writing volume (\eg emails, reports, and posts), two participants reported writing 1--3 hours per week, six reported 4--7 hours, two reported 8--14 hours, and two reported 15+ hours. 
Participants were generally experienced with prompting, with a mean self-rated expertise of $5.17$ ($SD = 0.58$) on a 7-point Likert scale (1 = Not at all, 7 = Very experienced). 
Each participant was compensated with a \$20 gift card per hour.
Appendix~\ref{appendix:participant_information} provides detailed participant information.\looseness=-1
% They are all familiar with the concepts involved in the tasks described below.

% \input{tables/tab-jeopardytasks}

\subsection{Study Design}
In this section, we describe the baseline and tasks.

\namedparagraph{Baseline}
The baseline resembles common conversational LLM interfaces such as ChatGPT.
In the baseline, users relied on NL prompting alone to generate outputs through a turn-by-turn conversation.
We used the same underlying model (\texttt{Qwen3-8B}) for output generation as in our system to ensure a fair comparison.
Example screenshots of the baseline can be found in Appendix~\ref{appendix:baseline_interface}.\looseness=-1

\namedparagraph{Tasks}
The study consisted of two parts: a Jeopardy evaluation and a free-form exploration.
The \textit{Jeopardy evaluation}, proposed by Gao et al.~\cite{gaoDataToneManagingAmbiguity2015a} and adopted by subsequent work~\cite{srinivasanOrkoFacilitatingMultimodal2018,srinivasanInChorusDesigningConsistent2020,srinivasanHowAskWhat2020,manamTaskLintAutomatedDetection2022,maAmbigChatInteractiveHierarchical2025,fengPromptMagicianInteractivePrompt2024}, is a well-established methodology for evaluating natural language interfaces in settings where users must explore and refine preferences.
Its core idea is to provide participants with a task and a concrete target output (a goal state), then ask them to use the system to reproduce the target as closely as possible.
This design avoids a common challenge in open-ended writing studies: if participants are given only a vague instruction (\eg ``write an email requesting a meeting'') without any specific goals, they may stop after receiving a random first draft, making it difficult to compare interfaces on iterative behavior.
\looseness=-1

For Jeopardy evaluation, we designed two writing tasks: an email task and an explanation task.
These tasks reflect a common writing scenario in which we must tailor content to different recipients or audiences~\cite{bellLanguageStyleAudience1984,bitzerRhetoricalSituation1968}.
For each task, we prepared three target versions that differed along multiple preference dimensions based on our taxonomy of the preference space (\tabref{tab:preference_taxonomy}) and prior work~\cite{leeAligningThousandsPreferences2024}, such as audience, tone, depth, and formatting, while preserving the same underlying intent and factual content.
\revision{Appendix~\ref{appendix:jeopardy_tasks}} shows the tasks and target versions.
To ensure fairness between conditions, we defined each target version using an explicit rubric of preference settings (\eg audience = manager, tone = formal, format = bullet-point plan) and then authored the final target texts to satisfy the rubric with the help of the baseline.
We iteratively refined targets to ensure that (1) successive versions were clearly distinguishable, (2) the intended preference differences were salient, and (3) each target was feasible to reproduce using either interface.
All targets were fixed prior to the study and held constant across conditions.
Supplemental materials include all target outputs presented to participants.
\looseness=-1

In addition to goal-directed Jeopardy tasks, we included a free-form exploration phase to capture more naturalistic behaviors that may not arise under explicit target matching.
This phase was designed to surface how participants appropriate the interfaces, what strategies they adopt for iterative refinement, and what breakdowns occur in self-directed writing workflows.
\revision{In addition, the Jeopardy task did not allow direct output editing to enable a controlled evaluation of prompting. However, because real-world writing often combines prompting with manual refinement, we allowed output editing at this stage to more realistically observe how \tool fit into users' practices.}
We employed a think-aloud protocol during this phase, followed by a semi-structured interview. Participants used the system for a diverse range of genres, including academic introductions, homework essays, slide scripts, argumentative essays, social media posts, blog posts, and creative writing such as poetry as shown in Appendix~\ref{appendix:participant_information}.
\looseness=-1

\subsection{Study Procedure}
The study began with informed consent and a demographics questionnaire.
Participants then completed the two parts of the study.
Each session lasted approximately 90 minutes.\looseness=-1

\namedparagraph{Jeopardy Evaluation}
In each condition, participants completed one Jeopardy task (either the email task or the explanation task).
At the start of each condition, participants received a brief tutorial and completed a short warm-up task to familiarize themselves with the interface (5 minutes).
For the assigned task, participants were asked to produce an output matching Version~1, then revise it to match Version~2, and finally revise it again to match Version~3.
We allotted 15 minutes for the Jeopardy task in each condition and allowed participants to generate as many drafts as needed within the time limit using NL prompting in the baseline or using \tool.
The order of systems and tasks was counterbalanced across participants.
% Participants indicated when they believed they had matched the target before proceeding to the next version.
After each condition, participants completed a short survey (2--3 minutes) including 11 items (6 derived from design goals in~\secref{sec:design_goals} and 5 from Wu et al.~\cite{wuAIChainsTransparent2022}) assessing user experience.\looseness=-1

\namedparagraph{Free-form Exploration}
After completing the two Jeopardy tasks, participants completed a time-boxed free-form writing session (15 minutes) with our prototype.
Participants worked on a self-chosen writing goal based on their recent real-life writing needs.
We encouraged participants to think aloud throughout.
After the free-form exploration, we conducted a short semi-structured interview (15 minutes) for their subjective feedback and interface-specific strategies and breakdowns (questions listed in Appendix~\secref{appendix:interview_questions}).\looseness=-1

\subsection{Quantitative Findings}
In this section, we report quantitative results from our analysis of participant event logs, task performance, and subjective ratings.
For statistical analysis, we used the Wilcoxon signed-rank test with Bonferroni correction due to the small sample size and the non-normal distribution of the data.\looseness=-1

\begin{table*}
\footnotesize
\centering
\caption{\formatcaption{Distribution of widget types across creation and adjustment.}{This summary includes the raw counts and percentages of widget types that participants created and adjusted across both the Jeopardy evaluation and free-form exploration tasks.}}
\label{tab:widget_distribution_combined}
\begin{tabular}{llccccc}
\toprule
\textbf{Measure} & \textbf{Task} & \textbf{Slider} & \textbf{Toggle} & \textbf{Dropdown} & \textbf{Stepper} & \textbf{Total} \\
\midrule
\multirow{3}{*}{Created}
& Jeopardy    & 181 (52.31\%) & 86 (24.86\%) & 54 (15.61\%) & 25 (7.23\%) & 346 \\
& Exploration & 96 (52.17\%)  & 44 (23.91\%) & 38 (20.65\%) & 6 (3.26\%)  & 184 \\
\cmidrule{2-7}
& \textbf{Total} & \textbf{277 (52.26\%)} & \textbf{130 (24.53\%)} & \textbf{92 (17.36\%)} & \textbf{31 (5.85\%)} & \textbf{530} \\
\midrule
\multirow{3}{*}{Adjusted}
& Jeopardy    & 385 (83.70\%) & 46 (10.00\%) & 21 (4.57\%) & 8 (1.74\%) & 460 \\
& Exploration & 274 (88.67\%) & 17 (5.50\%)  & 10 (3.24\%) & 8 (2.59\%) & 309 \\
\cmidrule{2-7}
& \textbf{Total} & \textbf{659 (85.70\%)} & \textbf{63 (8.19\%)} & \textbf{31 (4.03\%)} & \textbf{16 (2.08\%)} & \textbf{769} \\
\bottomrule
\end{tabular}
\Description{Table summarizing the distribution of widget types across creation and adjustment in the ``Jeopardy'' and ``Exploration'' tasks. For created widgets, 530 total widgets were created: sliders were most common at 277 (52.26\%), followed by toggles at 130 (24.53\%), dropdowns at 92 (17.36\%), and steppers at 31 (5.85\%). For adjusted widgets, 769 total adjustments were made: sliders dominated at 659 (85.70\%), followed by toggles at 63 (8.19\%), dropdowns at 31 (4.03\%), and steppers at 16 (2.08\%). This shows that sliders were by far the most frequently created and especially the most frequently adjusted widget type across both tasks.}
\end{table*}

% \begin{table} \footnotesize \centering \caption{\formatcaption{Distribution of created widgets by category.}{Summary includes the raw counts and percentages of widgets created by all participants across both the Jeopardy evaluation and free-form exploration tasks.}} \label{tab:widget_creation} \begin{tabular}{lccccc} \toprule \textbf{Type} & \textbf{Slider} & \textbf{Toggle} & \textbf{Dropdown} & \textbf{Stepper} & \textbf{Total}\\ \midrule Jeopardy & 181 (52.31\%) & 86 (24.86\%) & 54 (15.61\%) & 25 (7.23\%) & 346 \\ Exploration & 96 (52.17\%) & 44 (23.91\%) & 38 (20.65\%) & 6 (3.26\%) & 184 \\ \midrule \textbf{Total} & \textbf{277 (52.26\%)} & \textbf{130 (24.53\%)} & \textbf{92 (17.36\%)} & \textbf{31 (5.85\%)} & \textbf{530} \\ \bottomrule \end{tabular} \end{table}
\begin{table*}[t]
\footnotesize
\centering
\caption{\formatcaption{Distribution of widget types across preference categories.}{This summary details the frequency and categorization of widget types across preference categories (\tabref{tab:preference_taxonomy}) created by participants during both the Jeopardy evaluation and free-form exploration tasks.}}
\label{tab:combined_widget_categories}
\begin{tabular}{llcccccc}
\toprule
\textbf{Task} & \textbf{Type} & \textbf{Format \& Structure} & \textbf{Tone \& Style} & \textbf{Audience \& Persona} & \textbf{Length \& Conciseness} & \textbf{Content Constraints} & \textbf{Total} \\
\midrule
\multirow{5}{*}{Jeopardy}
 & Slider   & 30 (16.6\%) & \maxnumber{89 (49.2\%)} & 2 (1.1\%)   & 30 (16.6\%) & 30 (16.6\%) & 181 \\
 & Toggle   & 28 (32.6\%) & 2 (2.3\%)   & 0 (0.0\%)   & 1 (1.2\%)   & \maxnumber{55 (64.0\%)} & 86  \\
 & Dropdown & 0 (0.0\%)   & 4 (7.4\%)   & \maxnumber{26 (48.1\%)} & 1 (1.9\%)   & 23 (42.6\%) & 54  \\
 & Stepper  & 9 (36.0\%)  & 0 (0.0\%)   & 0 (0.0\%)   & \maxnumber{15 (60.0\%)} & 1 (4.0\%)   & 25  \\
 \cmidrule{2-8}
 & \textbf{Subtotal} & \textbf{67 (19.4\%)} & \textbf{95 (27.5\%)} & \textbf{28 (8.1\%)} & \textbf{47 (13.6\%)} & \textbf{109 (31.5\%)} & \textbf{346} \\
\midrule
\multirow{5}{*}{Exploration}
& Slider   & 16 (16.7\%) & \maxnumber{54 (56.2\%)} & 0 (0.0\%)   & 23 (24.0\%) & 3 (3.1\%)   & 96 \\
 & Toggle   & 6 (13.6\%)  & 8 (18.2\%)  & 2 (4.5\%)   & 0 (0.0\%)   & \maxnumber{28 (63.6\%)} & 44 \\
 & Dropdown & 11 (28.9\%) & 3 (7.9\%)   & \maxnumber{17 (44.7\%)} & 0 (0.0\%)   & 7 (18.4\%)  & 38 \\
 & Stepper  & 2 (33.3\%)  & 0 (0.0\%)   & 0 (0.0\%)   & 2 (33.3\%)  & 2 (33.3\%)  & 6  \\
 \cmidrule{2-8}
 & \textbf{Subtotal} & \textbf{35 (19.0\%)} & \textbf{65 (35.3\%)} & \textbf{19 (10.3\%)} & \textbf{25 (13.6\%)} & \textbf{40 (21.7\%)} & \textbf{184} \\
\midrule
\multicolumn{2}{l}{\textbf{Grand Total}} & \textbf{102 (19.2\%)} & \textbf{160 (30.2\%)} & \textbf{47 (8.9\%)} & \textbf{72 (13.6\%)} & \textbf{149 (28.1\%)} & \textbf{530} \\
\bottomrule
\end{tabular}
\Description{Table summarizing the distribution of created widget types across five preference categories in the ``Jeopardy'' and ``Exploration'' tasks. Across both tasks, 530 widgets were created. Tone and Style was the most common category at 160 widgets (30.2\%), followed by Content Constraints at 149 (28.1\%), Format and Structure at 102 (19.2\%), Length and Conciseness at 72 (13.6\%), and Audience and Persona at 47 (8.9\%). Sliders were used mainly for Tone and Style, toggles for Content Constraints, dropdowns for Audience and Persona, and steppers most often for Length and Conciseness. This pattern appeared in both tasks, though Exploration had a somewhat larger share of Tone and Style widgets than Jeopardy.}
\end{table*}

\namedparagraph{Widget Creation}
We first analyzed how participants created widgets across both tasks (\tabref{tab:widget_distribution_combined}). 
In the Jeopardy task, participants created 346 widgets, primarily sliders (\(N=181\), 52.31\%), followed by toggles (\(N=86\), 24.86\%), dropdowns (\(N=54\), 15.61\%), and steppers (\(N=25\), 7.23\%).
This distribution remained consistent in the Exploration task (\(N=184\), where sliders again led at 52.17\% (\(N=96\)), followed by toggles (\(N=44\), 23.91\%), dropdowns (\(N=38\), 20.65\%), and steppers (\(N=6\), 3.26\%).
Overall, across both tasks, participants created a total of 530 widgets, with sliders remaining the dominant choice (\(N=277\), 52.26\%).

We then categorized the selected preference expressions for each created widget based on the taxonomy of the preference space in~\tabref{tab:preference_taxonomy}. 
As shown in~\tabref{tab:combined_widget_categories}, the most prominent trend is the strong association between sliders and Tone \& Style. 
In both the Jeopardy ($49.2\%$) and Exploration ($56.2\%$) tasks, participants predominantly chose sliders to adjust stylistic elements, likely favoring the granular, continuous control they provide for subjective qualities.
In contrast, toggles were the primary choice for Content Constraints, accounting for over $60\%$ of toggle usage in both tasks. 
This suggests that participants viewed content constraints as binary decisions (either including or excluding specific information) which aligns naturally with the on/off functionality of toggles.
Dropdowns served as the primary instrument for defining Audience \& Persona, representing nearly half of all dropdown interactions ($48.1\%$ in Jeopardy and $44.7\%$ in Exploration). This indicates that users prefer selecting from a predefined list of distinct categories when characterizing an intended audience. 
Steppers, while used less frequently overall, were heavily utilized for Length \& Conciseness (\eg $60\%$ in the Jeopardy task) and Format \& Structure, where incremental adjustments to output size or structural complexity were required.
Overall, Tone \& Style ($30.2\%$) and Content Constraints ($28.1\%$) were the most frequently targeted categories for widget creation. 
% The consistency of these distributions across both tasks suggests that participants have intuitive mental models for which widgets are best suited for specific types of preference expressions.\looseness=-1

\namedparagraph{Widget Adjustment}
We then examined how frequently participants adjusted the values of these widgets after creation (\tabref{tab:widget_distribution_combined}). 
In the Jeopardy task, participants performed 460 configurations, with sliders accounting for the vast majority of interactions ($N=385$, 83.70\%), followed by toggles ($N=46$, 10.00\%). 
This trend was even more pronounced in the Exploration task, where sliders represented 88.67\% of all interactions ($N=274$). 
Across both tasks, steppers and dropdowns were configured the least frequently, totaling only 2.08\% and 4.03\%.
This indicates that while categorical widgets are typically used to establish stable constraints, sliders serve as the primary tool for iterative, granular refinement of the model's output.\looseness=-1

% This pattern suggests that participants primarily formulated preferences as continuous adjustments, using sliders as the dominant mechanism for iterative refinement.\looseness=-1
% This distribution suggests that participants most often expressed preferences through continuous controls, with categorical and binary controls used less often.
% This iterative cycle suggests that inline widgets lowered the cost of exploring prompt variations, shifting users from rewriting prompt text toward direct manipulation of extracted parameters.

\namedparagraph{Jeopardy Task Performance}
% In the Jeopardy evaluation tasks, we assessed quality by counting how many of the four target preferences for each version (\tabref{tab:jeopardy_tasks}) were precisely achieved, assigning 1 point per preference for a maximum score of 4.
In the Jeopardy evaluation, we assessed quality by evaluating whether each of the four target preferences for each version (Appendix~\ref{appendix:jeopardy_tasks}) was met. 
\revision{We used a two-rater validation process; one rater scored all outputs against predefined binary rubrics, while a second reviewed scores and flagged ambiguous cases for discussion.}
Each preference dimension was evaluated independently using a binary scoring system: 1 point if the criterion was clearly satisfied based on the overall impression and dominant characteristics of the text, 0 points otherwise. This yielded a maximum score of 4 points per version.
Quality scores were significantly higher in the \tool condition (\(M=3.61, SD=0.55\)) than in the baseline (\(M=3.31, SD=0.67\); \(W=60.50, p=.036^{*}\)). 
This suggests that \tool helped users achieve target preferences more precisely than NL prompting alone.\looseness=-1
% In contrast, differences in completion time (\(W=17.00, p=.092\)) and total iterations (\(W=31.00, p=.557\)) were not significant.

% When using \tool in the Jeopardy evaluation tasks, participants predominantly engaged in a tight ``tweak-and-test'' loop (\figref{fig:interaction_log}). 
% The most frequent transition was from widget configuration to content generation (\(M=8.42\), \(SD=4.87\) per participant), followed by transitions from prompt editing to widget configuration (\(M=4.83\), \(SD=1.91\)) and from generation back to further widget adjustment (\(M=4.58\), \(SD=4.15\)). 
% These patterns suggest that participants primarily treated widgets as lightweight controls for rapid iteration, repeatedly adjusting parameters and immediately testing their effects on the output.\looseness=-1

% \input{tables/tab-taskcompletion}
% \input{figures/fig-interactionlog}
\begin{figure*}
\centering
\includegraphics[width=\linewidth]{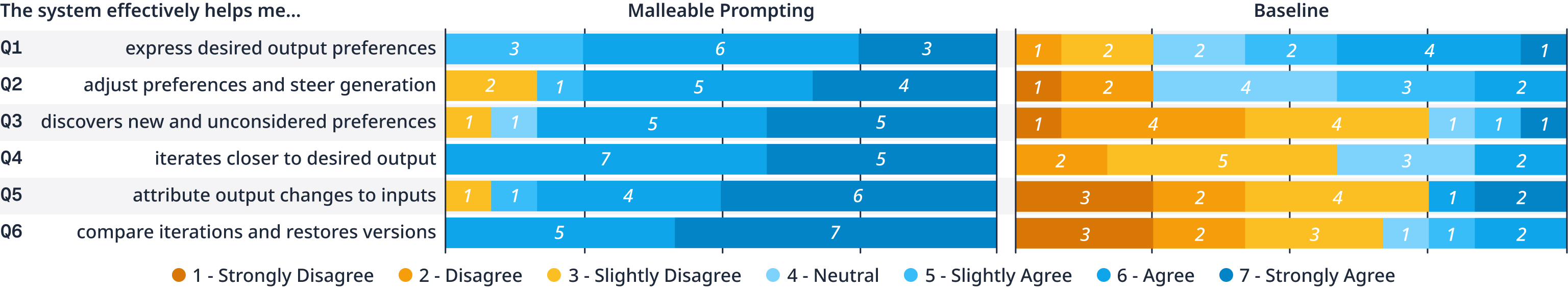}
\caption{\formatcaption{Perceived support for design goals.}{Participants' responses to a 7-point Likert-scale questionnaire, comparing our system against the baseline condition across both execution-related (Q1--Q4) and evaluation-related (Q5--Q6) measures.}}
\Description{``Malleable Prompting'' versus ``Baseline'' stacked horizontal Likert chart for six statements about whether the system helps users express and refine output preferences, rated from 1 (strongly disagree) to 7 (strongly agree). ``Malleable Prompting'' is strongly positive across all six questions, with responses concentrated in ratings 5 to 7. For Q1, all 12 responses are positive (3 slightly agree, 6 agree, 3 strongly agree). Q4 and Q6 are also entirely positive, with Q4 showing 7 slightly agree and 5 strongly agree, and Q6 showing 5 slightly agree and 7 strongly agree. Q3 and Q5 each include only one slightly disagree and one neutral response, with the rest positive. In contrast, the ``Baseline'' condition is much more mixed and often negative or neutral. Q1 and Q2 show some agreement but also include disagreement and neutral responses. Q3 through Q6 are dominated by ratings 1 to 4, with only a few agreement responses. The largest gaps favoring ``Malleable Prompting'' appear for discovering new preferences, iterating closer to the desired output, attributing output changes to inputs, and comparing or restoring versions.}
\label{fig:survey_goal}
\end{figure*}

\namedparagraph{Subjective Ratings}
% \chao{will update the data with Bonferroni correction}
As shown in \figref{fig:survey_goal}, participants rated malleable prompts significantly more favorably than the baseline across both execution- and evaluation-related measures. 
For executing preference changes, participants rated \tool and the baseline similarly in terms of conveying desired output preferences (\(M=6.00, SD=0.74\) vs. \(M=4.75, SD=1.54\); \(W=42.00, p=.141\)).
However, they rated \tool higher for adjusting and steering outputs (\(M=5.92, SD=1.08\) vs. \(M=4.00, SD=1.60\); \(W=55.00, p=.012^{*}\)), discovering new preferences (\(M=6.00, SD=1.28\) vs. \(M=3.08, SD=1.62\); \(W=55.00, p=.012^{*}\)), and iterating closer to desired output (\(M=6.42, SD=0.51\) vs. \(M=3.58, SD=1.31\); \(W=66.00, p=.006^{**}\)).
Participants also rated \tool significantly higher on evaluation-related abilities, including understanding cause--effect relationships between prompt changes and outputs (\(M=6.25, SD=0.97\) vs. \(M=3.25, SD=2.22\); \(W=55.00, p=.012^{*}\)) and comparing iterations (\(M=6.58, SD=0.51\) vs. \(M=3.08, SD=1.83\); \(W=78.00, p=.003^{**}\)).
This pattern was echoed in ratings of the overall experience: participants reported that \tool better helped them think through outputs (\(M=6.42, SD=1.00\) vs. \(M=3.67, SD=1.67\); \(W=55.00, p=.010^{**}\)), provided greater transparency (\(M=6.17, SD=0.83\) vs. \(M=3.08, SD=1.73\); \(W=66.00, p=.005^{**}\)), and gave them a stronger sense of control (\(M=6.25, SD=0.45\) vs. \(M=4.08, SD=1.62\); \(W=78.00, p=.002^{**}\)).
% \yiren{possibly add some analysis around users' decision making of choosing and using widgets during the free-form phase; Echo the taxonomy in this part of analysis as well}
\looseness=-1

\subsection{Qualitative Findings}

% \chao{link back to our claims about the two gulfs in intro. show our method helps}
We conducted qualitative analysis across all 12 participants' transcripts, following established thematic analysis protocols~\cite{braunUsingThematicAnalysis2006}. 
% The research group collaboratively analyzed the first three interview transcripts to develop an initial codebook. One researcher then applied this codebook to the remaining transcripts, while the group met over the course of a month to review coding decisions, discuss discrepancies, and refine the codebook iteratively. The finalized codebook was subsequently used to support thematic analysis and identify key themes.\looseness=-1

% For qualitative analysis of interview transcripts, we followed established open-coding protocols~\cite{braunUsingThematicAnalysis2006,scupinKJMethodTechnique1997}.
% The first three authors coded the transcripts, then discussed disagreement, reached a consensus, and created a consolidated codebook.
% This codebook was then used for thematic analysis to identify emerging topics from the interviews.\looseness=-1

% \yiren{TODO: trim + removing some confirmational quotes from the first two paragraphs; add more critical reflections for the third paragraph.}

% \namedparagraph{Enhancing Language Expressiveness with Widgets}
\namedparagraph{Fine-Grained Preference Control with Widgets}
We found that the widgets come into play when natural language lacks the expressive resolution to communicate fine-grained stylistic preferences. 
Participants (11/12) described moments where NL prompting alone was difficult to convey the degree of a desired attribute. 
As P12 mentioned in the baseline condition, \userquote{I want something shorter, but it gets way too short. I wanted 50\%, but it compressed to 20\%. I can only add more degree adverbs and try to move it back.} 
% P06 explained why widgets resolve this: \userquote{This function could also be achieved with degree adverbs, but there are only a few types. If this slider goes up, it can achieve more possibilities than degree adverbs.} 5/12 participants spontaneously drew this slider-as-adverb analogy.
In contrast, P06 commented that widgets \userquote{can achieve more possibilities than degree adverbs,} which was echoed by five other participants.
For example, P02 noted that fine-grained degree control \userquote{is something I can't describe well with language.} 
P02 also pointed out that \tool allowed them to perform more fine-grained edits over prompts: \userquote{Each prompt becomes very modular ... widgets let me adjust many small parts of the sentence. It makes my control over this prompt very granular.}
Our system's enrich feature also addressed a complementary problem: users often do not know which dimensions to control. All 12 participants valued this feature for augmenting a minimal prompt with rich specifications. 
P04 described this as \userquote{... it is always more difficult to create than to revise ... Starting with something to revise is much simpler.}
Users' varied language background also plays a role in this, as P06 described: \userquote{As a non-native English speaker ... I might not even know what kind of result I want myself, } which is echoed by three other participants.
On the other hand, we also observe some participants (5/12) specifying attributes in natural language by directly editing the input prompt when desired attributes are not present in widgets. 
\looseness=-1

% \namedparagraph{From Linear Chat History to Parametric Comparison}
\namedparagraph{Evaluating Outputs with Versioning and Attribution}
Participants reported that \tool's version control and attribution features helped them more effectively evaluate outputs \revision{(8/12). 
For example, P11 emphasized that the the version control feature helped them \userquote{break the linear interface and reuse previous prompts.}}
% Meanwhile, P10 described a click-and-compare workflow: \userquote{I can just click around and compare which one is most satisfying. I don't need to go back and forth.}
% Participants also found the attribution feature helpful for evaluation. 
P01 described that the attribution feature enabled them to evaluate trade-offs in tasks that require careful inspection, \userquote{I found two parts of my prompt were actually in a trade-off (emphasizing contributions vs. factual accuracy),} and the participant utilized the attribution feature to \userquote{guide} them to think about \userquote{which parts have impact on the results} and \userquote{which parts to change.}
% However, participants also preferred holistic evaluation over token-level probability-based visualization of attribute influence. 
% More specifically, several participants mentioned wanting coarser granularity in attribution (3/12).
However, some participants preferred coarser granularity in attribution (3/12) over word-level visualization of influence.
As P06 described: \userquote{I judge based on the overall text whether it matches the tone I want. I wouldn't judge through a single word.} 
Two participants suggested a feature that allows the system to automatically suggest widgets that are inferred ``in reverse'' from a selection within output: \userquote{select text here and have it tell me which widget can adjust that} (P12).
\looseness=-1

\namedparagraph{How Users Map Widgets to Preferences}
Participants reflected on their mental models of how to use \tool and different strategies for adopting widgets.
Sliders were preferred (11/12) for subjective and adjective-based attributes that vary by degree or ordinality such as tone, formality, warmth, and humor. 
Dropdowns mapped naturally to discrete categorical attributes (10/12) and surfaced hidden possibilities (8/12), as P07 reflected, \userquote{... it expanded the options to, for example, junior or more senior IS (Information Systems) scholars. These choice ranges, before it expanded them, I myself hadn't thought about the possibility of further adjustment.}
P04 found toggles \userquote{good for objective things like [whether to use] paragraphs or actionable insights (bullet points)}, also echoed by 6 other participants of being helpful for controlling whether to include bullet points, tables, or sections. However, 4 participants mentioned that in some cases, the toggling effect can be more easily achieved through the removal of certain expressions from the prompt.
\looseness=-1
% \userquote{(\tool is more helpful) when my content is fixed ... I only need to keep iterating whether the style meets my requirements ...} 
% Participants also identified scenarios where widgets offer both values and overhead in comparison with traditional chat interaction.

% HERE need to add the concrete themes+examples of what widgets the user liked or changed blablabla
% \yiren{todo}

\namedparagraph{Pros and Cons of Malleable Prompting}
% \chao{add some findings about the breakdowns (RQ3) of \tool? Actually, we have already mentioned some breakdowns earlier, maybe moving them to this part? we can also talk about \tool is not suitable for simple and determined tasks.}
Although widgets were generally preferred by participants, some participants also identified specific scenarios where either NL prompting alone or \tool offered more value.
Prompting alone was often preferred for simple, well-defined tasks (8/12), where parametric control may introduce unnecessary cognitive overhead.
Conversely, \tool was preferred for style-driven writing (11/12), tasks that can benefit from reusable templates (7/12), and tasks with many competing and/or undefined requirements (5/12). 
For example, P10 identified that scenarios where content is fixed, but tone, formality, and register require iteration, would benefit most from our method.
In addition, participants noted that their perception of the adjustment over sliders required calibration, especially for those attributes whose extremes are hard to imagine.
P02 described that for some complex attributes (\eg professional) they needed to \userquote{test the maximum first to find where the boundary is,} before intermediate values became meaningful.\looseness=-1

\section{Discussion}
\label{sec:discussion}
% Our findings show that \tool helps participants achieve target preferences more precisely and is perceived as more controllable and transparent than NL prompting alone.
Our findings provide empirical insights about how users approach \tool. Here, we discuss these insights and the implications around the two gulfs that motivated our work in \secref{sec:intro}.\looseness=-1
% 1) Sliders are heavily used post-creation and used most for adjustment related to ordinal attributes (e.g., tone and style), but natural language should be kept as the fallback for user-defined dimensions; 
% 2) enrichment feature is also preferred, as support for both ``defining'' preferences and ``adjusting'' attribute values is important; 
% 3) The attribution feature is liked for interpretation, but granularity should be semantic-aware. 
% 4) Additional designs should be introduced to handle potential conflicts between preferences.
\looseness=-1
% \yiren{TODO: revise the below sections following the above outline}

\subsection{Bridging the Gulf of Execution}
% \yiren{Why model-only generation? We address the prompt tuning phase, which is orthogonal to the text editing phase that DirectGPT~\cite{masson_directgpt_2024} and Textoshop~\cite{massonTextoshopInteractionsInspired2025} address. A full system could integrate both parametric input-side control complements output-side direct manipulation.}

\namedparagraph{When and Why \tool Helps}
% Our results echo the finding by \citet{zamfirescu-pereiraWhyJohnnyCant2023} that users struggle to edit over pre-existing prompts, which suggests that NL prompting lacks the expressive resolution required for fine-grained stylistic preferences --- an area where our approach excels. 
\tool helped mitigate users' challenges in specifying and editing preferences in natural language~ \cite{zamfirescu-pereiraWhyJohnnyCant2023}.
We also found that users developed emerging mental models of when to use which type of inline widgets. For example, sliders were found to be an effective design that allowed users to modulate existing ordinal attributes. However, when users needed specific attributes that had not yet been surfaced, they often resorted to inserting or appending natural language prompts. 
This suggests that while widget controls benefit the adjustment of preferences, natural language should still be offered as a fallback mechanism for user-defined attributes. Future systems should also consider allowing users to easily transit between these two control modalities.

\namedparagraph{Adaptive Widget Complexity for Different Task Types}
In terms of task type, users preferred NL prompting alone for simple, well-defined tasks, whereas \tool was preferred for style-driven writing and exploratory tasks involving competing or undefined requirements.
This was likely because the parametric control offered by our approach allowed them to manipulate specific dimensions of the output without the ambiguity or trial-and-error inherent in purely linguistic interfaces. At the same time, such control could also introduce overhead for simpler tasks. 
Similar to the concept of matching UI complexity to task complexity~\cite{richeAIInstrumentsEmbodyingPrompts2025}, our findings call for future designs to incorporate mechanisms to surface widgets adaptively, such as triggering when users began work on exploratory tasks rather than always being offered.\looseness=-1
% Another interesting observation is that \tool could be particularly beneficial for users with lower prompting proficiency, such as non-native speakers.
% Prior work indicates that users with language barriers face compounded challenges in AI-mediated writing, where the difficulty of articulating intent in a second language is exacerbated by the complexity of engineering effective prompts~\cite{xiaoDisplacedContributions2024}.
% \tool partially complements these skills by shifting the user's task from language production to parameter exploration.
% % Future work could further explore whether long term use of \tool could improve users prompting skills.
% % This has accessibility implications, users with less language fluency can still express preferences precisely, and connects to broader trends in malleable UI generation~\cite{caoGenerativeMalleableUser2025}.

\namedparagraph{Enrichment as Preference Discovery}
Beyond adjusting existing attributes, participants also valued \tool for discovering what attributes to control in the first place. 
Past research~\cite{slovicConstructionPreference1995,zamfirescu-pereiraWhyJohnnyCant2023} has suggested that it is easier for users to construct or ``envision''~\cite{subramonyamBridgingGulfEnvisioning2024} preferences through interactions. 
In \tool, the action space of widgets (e.g., range of sliders, options in dropdowns) was viewed as inspiration to scaffold this construction because they offered concrete alternatives to help users identify preferences they did not know they had.
To enhance this process, future work could explore additional implicit signals such as users' natural language edits to aid preference discovery and provide attribute suggestions on-the-fly.
\looseness=-1

\subsection{Bridging the Gulf of Evaluation}
% \todo{...}

% \chao{mention the gulf of envision?}

\namedparagraph{From Linear to Modular Prompt Iteration}
In standard chat interfaces, iterations are often organized chronologically, which makes it challenging to compare outputs or attribute differences to specific prompt changes~\cite{geroSupportingSensemakingLarge2024,zhangNavigatingFogHow2025}. 
Our system restructures this by organizing iterations around changes in configuration of parameters rather than chat turns. This design was valued by participants, reflected through their preferences for the version control feature. 
Additionally, by making the mapping between the changes in widget values and outputs explicit, \tool turns evaluation from traditional holistic judgment to a more structured and comparative approach~\cite{normanCognitiveEngineering1986a}.
%  Attribution Granularity
% The attribution feature was also perceived as helpful for interpreting the impact of widgets. 
We also found that participants preferred coarser granularity in attribution over word-level visualization of influence. Attribution granularity can be designed as attribute-type-aware, such as using sentence-level attribution for tone and style or word-level attribution for localized constraints.\looseness=-1 
% This also points to a broader design implication that interpretation support should be provided for users to compare between different versions of prompts throughout edits. 

\namedparagraph{Calibrating Perception Through Boundary Probing}
Our findings also suggest that users naturally adopt a calibration strategy enabled by continuous controls.
P02's mention of \userquote{test the maximum first} reveals that continuous controls require aligning their perception of the slider's range before intermediate values become meaningful.
This echos the challenge of ``gulf of envision''~\cite{subramonyamBridgingGulfEnvisioning2024} when interacting with LLMs.
Future designs can consider showing example outputs at sliders' boundary values, or providing previews to help users align their perception of the slider's range.

\subsection{Broader Applications}
\label{sec:broader_applications}
Here, we discuss the potential application of \tool beyond the example system in~\secref{sec:system_design} and use cases from our study.
% \subsubsection{Integrating into a text editor?}

%JRz - is the hook here that we can make additional inferences from the text the user provides / interactions with the editor?
\namedparagraph{Text Editor}
\tool enables prompt-side parametric control, which can be combined with output-side direct manipulation (\eg DirectGPT~\cite{masson_directgpt_2024}) to be integrated into a working text editor for real-life use.
One participant also explicitly mentioned that a ``reverse widget'' can be useful in the case where the user wants to indicate a specific output area where refinement is needed, and the system can thus infer and create an appropriate widget that only impacts the target range.
Additionally, the range can also be user-defined through manual selection.
% \yiren{todo}

% \namedparagraph{Integrating malleable prompting for workflow usability}
\namedparagraph{Prompt Library}
\tool also points toward reusable prompt templates with built-in controls. 
Several participants saw value in templates that could be shared across users, with personalization supported through user-specific slider calibrations, or reused across tasks with different attribute configurations.
This direction connects to prior work on prompt reuse challenges~\cite{geroSupportingSensemakingLarge2024} and composable prompt workspaces~\cite{aminComposablePromptingWorkspaces2025}. 
Designing and implementing such libraries would require a structured template schema that encodes attribute types, widget types, and valid ranges or options. 
% A key design challenge is how to separate reusable task specifications (e.g., prompt instruction for generating a professional email) from customizable user preferences (e.g., tone and target audience).
% \chao{we actually already have this part in Appendix~\ref{appendix:segmenting_prompts}}
Additional features such as automatic recommendation and adaptation can also be implemented to enhance re-usability of such templates for different use cases.

% \subsubsection{Malleable agentic workflow?}
% \namedparagraph{Steering with widgets beyond writing styles}
\namedparagraph{Agentic Workflow}
The steering mechanism can potentially bootstrap beyond style-driven writing to broader agentic workflows. For example, controllable parameters could help steer code generation, data wrangling, information gathering, tool-use strategies, or brainstorming processes.
In these settings, widgets may capture not only stylistic preferences but also behavioral ones, such as exploration depth, level of initiative, or degree of constraint in tool use. 
Supporting these applications would require validating and extending our preference taxonomy~\cite{leeAligningThousandsPreferences2024} for domains beyond content generation.\looseness=-1

% \vspace{-6pt}

%JRz: People create chains of prompts to accomplish complex tasks -- could we use this approach to make changes to series of prompts all at once

\subsection{Limitations and Future Work}

Our work has several limitations that could be addressed in future work.
First, 
% our decoding algorithm involves two forward passes for each controlled attribute, which may introduce computational overhead. We mitigated this extra latency through batched inference and KV-cache reuse. A detailed analysis of the computational overhead is provided in Appendix~\ref{appendix:computation_overhead}.
% Second, 
our study was conducted with a smaller user sample of 12 participants and findings might not generalize to users with broader background and used to work with other writing scenarios. 
Second, our analysis did not consider participants' expertise level in writing and prompt engineering, which may affect the participants' strategy and breakdowns when using \tool. 
Third, our Jeopardy tasks focused on emails and explanations, while more open-ended creative tasks where preferences are less decomposable, such as poetry or fiction writing, should be considered.\looseness=-1
% Additionally, structured widgets also risk anchoring users to extracted dimensions, leading to homogenized outputs~\cite{wanDiverseAIPersonas2025,shinInterrogatingDesignHomogenization2026}. 
% The enrichment feature may frame all subsequent exploration around suggested dimensions, potentially limiting serendipitous discovery. 
% Future work should study whether initial suggestions constrain the preference space users explore.
% Also, our study did not consider longitudinal use scenarios, in which the number of widgets may increase significantly with continued use. In such cases, the stability of the widgets' effects and the cost of calibration should be examined more carefully.
% \looseness=-1

\revision{
Technically, our steering method provides an operational approximation for context-based control rather than globally grounded, fully disentangled semantic parameters. 
Subjective attributes such as ``funny'' and ``professional'' are context-dependent and not able to be fully disentangled. 
Accordingly, sliders should be interpreted as contextual control handles: although they do not represent universal semantic dimensions, they remain valuable for controlling preferences and steering models, as shown in our study.
Similarly, our attribution method provides a local, post-hoc, probability-based signal intended as a perceptual handle rather than a precise causal explanation of the model's reasoning. 
Our technical evaluation should also be interpreted as a task- and model-specific validation for our system design rather than a definitive measure of semantic alignment or the causal faithfulness of this token-control method.
Future work could explore better semantic disentanglement, causal explanations of widget effects, and more generalizable evaluations across models, settings, and tasks.
\looseness=-1
}

\section{Conclusion}
We introduce \tool, an interaction technique that expands preference expression from natural language prompts to in-line widgets, enabled by an LLM decoding method that steers and attributes each widget's influence on the output.
Our user study (N=12) showed that \tool helped users achieve target preferences more precisely and was perceived as more controllable and transparent than natural language prompting alone. 
We provide insights on how \tool bridges the gulfs of execution and evaluation that exist in chat-based prompting, and offer design suggestions for how future systems can combine GUI widgets and NL prompting as an approach for more precise and transparent human-AI collaboration.
\begin{acks}
\revision{
This material is based upon work supported by the National Science Foundation under Grant No. 2119589. Any opinions, findings, and conclusions or recommendations expressed in this material are those of the author(s) and do not necessarily reflect the views of the National Science Foundation.
This project was made possible in part by the Institute of Museum and Library Services RE-252329-OLS-22.
Additionally, results presented in this paper were obtained using CloudBank~\cite{norman2021cloudbank}, which is supported by the National Science Foundation under award No. 1925001.
}
\end{acks}

\balance
\bibliographystyle{ACM-Reference-Format}
\bibliography{reference}

\clearpage
\appendix
\section{Computation Overhead Analysis}
\label{appendix:computation_overhead}

We analyze the additional computation introduced by the two technical components of \tool relative to standard single-pass decoding, which requires exactly $N$ forward passes to generate an output of $N$ tokens.

\paragraph{Prompt Modularization}
At each decoding step $i$, computing the modulated distribution (Equation~3) requires evaluating $F_{a,P_0}(x_i)$ for each $a \in \mathcal{A}_{\text{mod}}$, including binary attributes in $\mathcal{A}^{\text{bin}}$ and continuous attributes in $\mathcal{A}^{\text{cont}}$. This process involves two forward passes per attribute: one with attribute $a$ present, one without.
In addition, one forward pass is required for the substituted base $p(x_i \mid x_{<i}, P_{\text{sub}})$.

However, the pass for $p(x_i \mid x_{<i}, \varnothing, P_0)$ is shared across all attributes in $\mathcal{A}_{\text{mod}}$, so the total per-step cost is $|\mathcal{A}_{\text{mod}}| + 2$ forward passes: one for $P_{\text{sub}}$, one for $\varnothing$, and one for each $a \in \mathcal{A}_{\text{mod}}$.
Over $N$ decoding steps, prompt modularization requires:
\begin{equation}
C_{\text{mod}} = (|\mathcal{A}_{\text{mod}}| + 2) \cdot N \text{ forward passes.}
\end{equation}

\paragraph{Prompt Attribution}
Computing the attribution $\phi_a(x_i)$ (Equation~4) requires evaluating $p_{\boldsymbol{\lambda}_{\setminus a}}$, the modulated distribution with attribute $a$'s weight set to zero.
Since the logits from every context pass are already in memory from the previous prompt modularization, this leave-one-out distribution is obtained by arithmetically recombining existing logits, with no additional forward passes.
Attributes in $\mathcal{A}_{\text{sub}}$ are handled by prompting an LLM to locate verbatim output spans, incurring one API call per attribute in $\mathcal{A}_{\text{sub}}$.

\paragraph{Total Overhead}
The total number of forward passes required for one generation-plus-attribution cycle is:
\begin{equation}
C_{\text{total}} = C_{\text{mod}} + C_{\text{attr}} = (|\mathcal{A}_{\text{mod}}| + 2) \cdot N.
\end{equation}
The overhead relative to standard single-pass decoding is therefore a factor of $(|\mathcal{A}_{\text{mod}}| + 2)$. To mitigate the practical wall-clock cost, we incorporate two acceleration techniques: 

a) \textbf{Batched inference}: We process all $|\mathcal{A}_{\text{mod}}| + 2$ sequences in a single batched forward call per step, making full use of GPU parallelism so the factor-of-$(|\mathcal{A}_{\text{mod}}| + 2)$ overhead does not multiply latency directly.

b) \textbf{Incremental KV-cache decoding}: To exploit the fact that all sequences append the same steered token at each step, their KV caches therefore remain synchronized throughout generation, so the model only needs to process one new token per sequence per step rather than reattending to the full context.

\section{Prompts}
\label{appendix:technical_details}
This section presents the prompt we used to instruct the LLM to enrich, parse, and segment prompts; generate options; and locate strings in the output for discrete attributes. 

\subsection{Enriching Prompts}
\begin{myfancybox}
\footnotesize

\noindent\textbf{System Prompt:}

{\ttfamily\raggedright\obeylines
You enrich a given content creation prompt by inferring missing user preferences.
\bigskip
Rules:
- Infer at most three missing preferences that the user did NOT state but are relevant and useful for the given task in the prompt.
- Use the below common preferences as a guide, but do not limit yourself to just these.
- Integrate the inferred preferences naturally into the original prompt.
- Ensure the enriched prompt remains clear and coherent.
\bigskip
Preference seed types (examples, not exhaustive):
- Formality: formal, informal
- Clarity: simple language, complex language
- Conciseness: concise, verbose/lengthy, clear, non-repetitive
- Vividness: use rhetorical devices (metaphors, personification, similes, hyperboles, irony, parallelism)
- Format: breadth-first, depth-first, step-by-step, consistency, deductive, inductive, parallelism, bullet points, narrative, satisfy constraints, interactive, support stances
- Tone: agreeable, sympathetic, cooperative, modest, altruistic, appreciative, forgiving, generous, kind, friendly, polite, funny, gregariousness, assertiveness, friendliness, extraversion, excitement-seeking, activity-level, cheerfulness, energetic, enthusiastic, talkable, anxiety, introspection/private self-consciousness, neuroticism, aloof, anger, depression, immoderation, vulnerability, self-pitying, self-beneficial, tense, touchy, unstable, worrying, emotionality, intellect, adventurousness, intellectual openness, liberalism, openness to experience, curious, authoritative, persuasive, professional, playful
- Depth: general topic, specific topic, nuanced insights
- Creativity: artistic, insightful, original, imaginative, novel, explorative creativity
- Efficiency: efficient, achievement-striving, self-discipline, self-contained, contain rich info
- Practicality: practical, use supporting materials, tailored examples and anecdotes, empowering actionable insights
- Audience: basic, novice, intermediate, advanced, expert, general public, for my boss, for a peer collaborator, for a family member, for a customer
}

\noindent\dotfill

\bigskip
\noindent\textbf{Context Prompt:}

{\ttfamily\raggedright\obeylines
Base prompt: \colorbox{grey-bg}{<insert the prompt>}
\bigskip
Infer missing preferences and return only the enriched prompt.
}

\end{myfancybox}

\subsection{Parsing Prompts}
\begin{myfancybox}
\footnotesize

\noindent\textbf{System Prompt:}

{\ttfamily\raggedright\obeylines
You extract controllable preferences from a given prompt for content creation.
\bigskip
Preference Types
\bigskip
1. Binary (on/off toggles)
For features that can be enabled/disabled.
- "Include examples"
- "Using bullet points"
\bigskip
2. Continuous (scalar intensity)
Use continuous for preferences that can vary along a spectrum (degree/strength/level). These are often adjectives or adverbs that describe how something should be (e.g., degree, mood, quality, price).
- "formal", "humorous", "casual", "friendly", "professional"
- "concise", "brief", "thorough", "verbose", "beginner-friendly"
- "cheap", "budget", "expensive", "luxury", "premium"
\bigskip
3. Categorical (discrete choices)
Use for mutually exclusive selections from a known/implicit set. These are often nouns or labels (audience, domain, locale, evaluation criteria, etc.).
- "for my boss"
- "in Manhattan, NY"
- "Chinese cuisine"
- "compare price and hardware"
- "explain like I'm five"
\bigskip
4. Numeric (quantities)
For counts and durations.
- "one-week extension"
- "3 paragraphs"
\bigskip
Output Fields
- id: lowercase-hyphenated identifier
- type: "binary" | "continuous" | "categorical" | "numeric"
- anchorText: EXACT substring from input (verbatim match required)
\bigskip
Preference extraction rules
- Returning an empty list is valid if no clear preferences are present.
- Do not invent preferences or infer beyond the text.
- Do not treat content type/genre (e.g., "blog post", "tweet", "email") as a preference.
}

\noindent\dotfill

\bigskip
\noindent\textbf{Context Prompt:}

{\ttfamily\raggedright\obeylines
Analyze this prompt and extract all controllable attributes: \colorbox{grey-bg}{<insert the prompt>}
}

\end{myfancybox}

\subsection{Segmenting Prompts}
\label{appendix:segmenting_prompts}
\begin{myfancybox}
\footnotesize

\noindent\textbf{System Prompt:}

{\ttfamily\raggedright\obeylines
You are a prompt segmentation assistant that separates a prompt into its core task and context prompts.
\bigskip
Given an original prompt and lists of attributes, you must:
\bigskip
1. Extract the BASE PROMPT: Remove all anchorText related to the attributes from the original prompt, keeping only the core task. The base prompt should be a neutral instruction without any context or binary prompt phrases.
2. Generate CONTEXT PROMPTS: For each attribute that needs to be turned into a context prompt, turn the provided anchorText into a standalone sentence. Preserve the attribute id.
\bigskip
Important rules:
- The base prompt should still be a complete, grammatical sentence
- Context prompts should start with "Be", "Keep", "Use", or similar directive
\bigskip
Example:
Original: "Write a concise and formal email to my boss asking for a one-week extension in two paragraphs, but keep the closing friendly. No bullet point."
Attributes to extract as context prompts: 
[\{ "id": "concise", "anchorText": "concise" \},
\{ "id": "formal", "anchorText": "formal" \},
\{ "id": "friendly", "anchorText": "friendly" \}]
\bigskip
Output:
- basePrompt: "Write an email to my boss asking for a one-week extension."
- contextPrompts: 
[\{ "id": "concise", "prompt": "Be concise" \},
\{ "id": "formal", "prompt": "Be formal" \},
\{ "id": "friendly", "prompt": "Keep the closing friendly" \}]
}

\noindent\dotfill

\bigskip
\noindent\textbf{Context Prompt:}

{\ttfamily\raggedright\obeylines
Original prompt: \colorbox{grey-bg}{<insert the prompt>}
\bigskip
Attributes to extract as context prompts: \colorbox{grey-bg}{<insert the attributes and their anchored text>}
\bigskip
Segment this prompt into a base prompt and context prompts.
}

\end{myfancybox}

\subsection{Generating Options}
\begin{myfancybox}
\footnotesize

\noindent\textbf{System Prompt:}

{\ttfamily\raggedright\obeylines
You create categorical options for a selected phrase in a user prompt.
\bigskip
Rules:
- Provide 3-5 mutually exclusive options.
- "value" is a short label (1-3 words).
- "text" is the full phrase that could replace the anchorText in the original prompt.
- Always include the selected text verbatim as one of the options.
- Keep each option aligned to the original prompt context.
- NEVER generate synonyms or near-synonyms of the selected text. Each option must lead the prompt in a meaningfully different direction.
- Think about what ROLE the selected text plays in the prompt, then offer diverse alternatives that fit that same role but change the prompt's focus.
}

\noindent\dotfill

\bigskip
\noindent\textbf{Context Prompt:}

{\ttfamily\raggedright\obeylines
Prompt: \colorbox{grey-bg}{<insert the prompt>}
\bigskip
Selected text: \colorbox{grey-bg}{<insert the anchored text>}
\bigskip
Generate categorical options that replace the selected text.
}

\end{myfancybox}

\subsection{Locating Strings}
\begin{myfancybox}
\footnotesize

\noindent\textbf{System Prompt:}

{\ttfamily\raggedright\obeylines
You are given a pair of prompt and output text. Your task is to locate exact, verbatim spans in the output text that are determined by the given portion in the prompt.
\bigskip
Rules:
- Only return substrings that appear exactly in the text.
- If there is no relevant span, return an empty list for that portion.
- Do not paraphrase or infer missing text.
}

\noindent\dotfill

\bigskip
\noindent\textbf{Context Prompt:}

{\ttfamily\raggedright\obeylines
Prompt: \colorbox{grey-bg}{<insert the prompt>}
\bigskip
Output text: \colorbox{grey-bg}{<insert the output text>}
\bigskip
Portion from the prompt (locate the exact output spans determined by this portion): \colorbox{grey-bg}{<insert the anchored text>}
}

\end{myfancybox}

% \section{Prompts for LLM-as-Judge}
% \label{appendix:llm_as_judge}

% \todo{...}

\section{Example Outputs for Different $\lambda$ Values}
\label{appendix:examples}
This section presents example outputs generated with different $\lambda$ for continuous attributes. The examples in \cref{tab:lambda_examples_combined} show how changing $\lambda$ alters the degree to which a selected attribute shapes the generated text.
% \looseness=-1

\begin{table*}[tb!]
\footnotesize
\centering
\caption{\formatcaption{Example effects of $\lambda$ across continuous attributes.}{Outputs produced from the same prompts while varying the strength of target attributes during decoding. The top example shows the attribute \textit{enthusiastic} for prompt ``\textit{Write a short post announcing a campus sustainability event in an enthusiastic tone},'' and the bottom example shows \textit{playful} for prompt ``\textit{Explain what a large language model is and what it can/cannot do in a playful way.}'' In both cases, larger $\lambda$ values produce a stronger realization of the target attribute while preserving the underlying content. Highlighted spans indicate wording influenced by the target attribute.}}
\label{tab:lambda_examples_combined}
\begin{tblr}{
  colspec = {l|c|X},
  hline{1,Z} = {0.8pt, solid},
  row{1} = {font=\bfseries},
  cells = {valign=t},
}
\textbf{Attribute} & \textbf{$\lambda$} & \textbf{Example output excerpt} \\
\hline
\SetCell[r=4]{l} \textit{enthusiastic}
& 0 &
Join us for Campus Sustainability Day on April 22nd. The event will include workshops, activities, and opportunities to learn about eco-friendly practices and environmental responsibility. Attend to learn how you can support a more sustainable campus community. \\
& 1 &
\influence{Join us} for Campus Sustainability Day on April 22nd! The event will feature workshops, activities, and opportunities to learn about eco-friendly practices and environmental responsibility. \influence{Come be part of} a more sustainable campus community. \\
& 2 &
\influence{Get ready to make a difference} at Campus Sustainability Day on April 22nd! Join us for engaging workshops, hands-on activities, and inspiring opportunities to learn about eco-friendly practices and environmental responsibility. \influence{Come celebrate and help build} a more sustainable campus community! \\
& 3 &
\influence{Get ready to celebrate, connect, and make a difference} at Campus Sustainability Day on April 22nd! Join us for \influence{exciting} workshops, hands-on activities, and inspiring opportunities to explore eco-friendly practices and environmental responsibility. \influence{Let’s come together and show our love for a greener future} by building a more sustainable campus community! \\
\hline
\SetCell[r=4]{l} \textit{playful}
& 0 &
A large language model is an advanced type of artificial intelligence that is trained on vast amounts of text data to understand and generate human-like text. It can perform tasks such as answering questions, writing stories, coding, and translating languages, but it cannot truly understand or possess consciousness. Additionally, it may sometimes produce incorrect or biased information, so its outputs should be critically evaluated. \\
& 1 &
A large language model is like a \influence{super-smart digital brain} that's been trained on tons of text to understand and generate human-like language---think of it as a \influence{chatbot with a PhD}! It can write stories, answer questions, translate languages, and even help with coding, but it can't truly understand emotions or have real-world experiences like humans do. Basically, it's a helpful tool with a lot of knowledge, but it's still just a bunch of words and patterns---it doesn't have a soul, just a lot of data! \\
& 2 &
Imagine a \influence{super-smart, never-sleeping librarian} who knows everything\ldots well, almost everything. A large language model is like that librarian, trained on tons of books, chats, and stories, so it can chat, write, explain things, and even help you brainstorm---but it can't actually read a book or see the world like you do. Basically, it's your \influence{friendly AI sidekick} who's great at talking, but still learning how to feel the vibe of a conversation! \\
& 3 &
Imagine a \influence{super-smart gossip squad} that's been training on every conversation, book, movie, and meme ever made---it's like having a \influence{party animal who's also a genius}! This is a Large Language Model (LLM), and it can chat, write stories, answer questions, and even help you brainstorm your \influence{next big idea}. But don't expect it to fold your laundry or tell you what your cat is thinking---it's great at words, but still learning how to do life hacks! \\
\end{tblr}
\Description{A three-column table comparing the effect of increasing decoding strength $\lambda$ across two continuous attributes, ``enthusiastic'' and ``playful.'' The first column lists the attribute, the second lists $\lambda$ values 0 through 3, and the third shows example output excerpts. For \textit{enthusiastic}, the prompt is about announcing Campus Sustainability Day. At $\lambda=0$, the announcement is neutral and informational. As $\lambda$ increases to 1, 2, and 3, the same event description becomes progressively more energetic, with highlighted phrases such as ``Join us,'' ``Get ready to make a difference,'' and ``Let’s come together and show our love for a greener future.'' For \textit{playful}, the prompt asks for an explanation of what a large language model is and what it can or cannot do. At $\lambda=0$, the explanation is formal and straightforward. As $\lambda$ increases, the explanation becomes progressively more whimsical and exaggerated, using highlighted phrases such as ``super-smart digital brain,'' ``super-smart, never-sleeping librarian,'' and ``super-smart gossip squad.'' Overall, the table shows that larger $\lambda$ values produce a stronger realization of the target attribute while preserving the underlying content of each prompt.}
\end{table*}

\revision{
\section{Technical Evaluation of Contextual Steering}
\label{appendix:cos_eval}

\begin{table*}[!h]
\footnotesize
\centering
\caption{\revision{\formatcaption{Slider-responsiveness regression.}{Least-squares slopes $\beta$ of judge score on intensity level (1-7). $\beta$ is measured in judge-score points per slider step.}}}
\label{tab:cos_beta}
\begin{tabular}{lcccc}
\toprule
\textbf{Scope} & \textbf{$\beta$ Baseline [95\% CI]} & \textbf{$\beta$ Ours [95\% CI]} & \textbf{$\Delta\beta$ [95\% CI]} & \textbf{p} \\
\midrule
Overall & 0.306 [0.278, 0.335] & 0.576 [0.546, 0.607] & +0.270 [0.228, 0.312] & $<.001^{***}$ \\
$k{=}1$ & 0.450 [0.382, 0.518] & 0.741 [0.673, 0.809] & +0.291 [0.195, 0.388] & $<.001^{***}$ \\
$k{=}2$ & 0.306 [0.255, 0.357] & 0.711 [0.661, 0.760] & +0.404 [0.333, 0.476] & $<.001^{***}$ \\
$k{=}3$ & 0.259 [0.220, 0.298] & 0.432 [0.388, 0.476] & +0.173 [0.115, 0.232] & $<.001^{***}$ \\
\bottomrule
\end{tabular}
\end{table*}

\namedparagraph{Methods}
We randomly sampled 100 prompts from the \texttt{WildChat-WC\textsubscript{Wr}} dataset \cite{mysorePrototypicalHumanAICollaboration2025a} and 35 style attributes from the preference corpus from~\citet{leeAligningThousandsPreferences2024}. We test the method under 30 steering settings: 10 single-attribute settings ($k{=}1$), 10 two-attribute combinations ($k{=}2$), and 10 three-attribute combinations ($k{=}3$), where $k$ denotes the number of attributes controlled simultaneously.

For each setting, we scan through the intensity of every active attribute over a 7-point scale and generated outputs under two conditions: 
1) In the prompt-based scaling baseline, the intensity was expressed as an explicit numeric instruction prepended to the prompt (\eg ``\textit{Apply the following writing style settings: formality: 2/7}'')~\cite{massonTextoshopInteractionsInspired2025}, and outputs were generated with standard decoding; 
2) In our method, each intensity level $s \in \{1, \dots, 7\}$ was mapped linearly to a steering coefficient spanning $\lambda \in [-10, 10]$, with the midpoint ($s{=}4$) recovers the unsteered distribution ($\lambda{=}0$), lower levels suppress the attribute, and higher levels amplify it.
Both conditions used the same underlying model (\texttt{Qwen3-8B}) and decoding parameters (\textit{temperature}${=}1$, \textit{top-p}${=}0$), yielding 4,200 paired observations.
An LLM judge rated the perceived intensity of each active attribute in each output on the same 1-7 scale; the judging protocol was validated against human ratings as described in \secref{sec:prompt_steering}.\looseness=-1

\namedparagraph{Results}
We report results from the \texttt{GPT-5.4} judge that our method achieving a steeper intensity-score linear regression slope on 29/35 dimensions ($p < .001^{***}$).
More specifically, across all 35 dimensions, our method roughly doubles the steering effect over the prompt-only baseline ($\beta = 0.576$, 95\% CI $[0.546, 0.607]$ vs.\ $\beta = 0.306$, 95\% CI $[0.278, 0.335]$). We were able to separate the effect with a significant interaction model ($\mathrm{score} \sim s \times \mathrm{condition}$; $\Delta\beta = +0.270$, 95\% CI $[0.228, 0.312]$, $p < .001^{***}$). As shown in \tabref{tab:cos_beta}, the better steering effect holds even at when multiple attributes are steered at once ($\Delta\beta = +0.29$, $+0.40$, and $+0.17$ for $k = 1$, $2$, and $3$, respectively; all conditions show $p < .001^{***}$). When examined by individual dimensions, our method yields a steeper steering intensity slope on 30 out of 35 dimensions. Adopting contextual steering enables stronger effect over dimensions that prompting has close-to-none effect, such as Tense ($\beta = 0.10 \rightarrow 0.99$), Immoderation ($0.27 \rightarrow 1.06$), Unstable ($0.09 \rightarrow 0.87$), Anger ($0.17 \rightarrow 0.81$), and Forgiving ($0.23 \rightarrow 0.84$). Conversely, there are 5 out of 35 dimensions our method were not able to steer more effectively than prompting alone: Agreeable ($\beta = 0.32 \rightarrow 0.04$), Friendly ($0.34 \rightarrow 0.10$), Generous ($0.31 \rightarrow -0.08$), Warmth ($0.40 \rightarrow 0.25$), and Active ($0.51 \rightarrow 0.48$).
}

\revision{
\section{Faithfulness of Attribution Highlights}
\label{appendix:attribution_eval}

\begin{table}[!h]
\footnotesize
\centering
\caption{\revision{\formatcaption{Attribution faithfulness and specificity.}{Directional sign-agreement between a token's displayed attribution label and the observed direction of its probability change under teacher forcing. Higher is more faithful. The neutral band (H1) and off-diagonal (H2) rows serve as controls that should sit near chance. $n$ is the number of prompt-attribute settings aggregated.}}}
\label{tab:attribution}
\begin{tabular}{lcc}
\toprule
\textbf{Condition} & \textbf{Sign-agreement [95\% CI]} & \textbf{$n$} \\
\midrule
\multicolumn{3}{l}{\textbf{H1: Faithfulness}} \\
\quad Highlighted            & 81.4\% [81.0, 81.7] & 598 \\
\quad Neutral (control)      & 67.2\% [66.4, 68.0] & 598 \\
\cmidrule(lr){1-3}
\quad \textit{by label polarity} & & \\
\quad\quad Positive          & 85.1\% [84.8, 85.4] & 598 \\
\quad\quad Negative          & 58.1\% [57.0, 59.2] & 580 \\
\cmidrule(lr){1-3}
\quad \textit{by no.\ of active attributes} & & \\
\quad\quad $k{=}1$           & 83.2\% [82.5, 84.0] &  98 \\
\quad\quad $k{=}2$           & 81.3\% [80.8, 81.8] & 200 \\
\quad\quad $k{=}3$           & 80.8\% [80.4, 81.2] & 300 \\
\midrule
\multicolumn{3}{l}{\textbf{H2: Specificity}} \\
\quad Diagonal (own)         & 81.0\% [80.7, 81.3] & 500 \\
\quad Off-diagonal (other, control) & 60.8\% [60.4, 61.3] & 800 \\
\bottomrule
\end{tabular}
\end{table}

\namedparagraph{Methods}
To validate that the attribution highlights described in \secref{sec:technical_framework} faithfully reflect a slider's effect, we evaluate two properties: faithfulness, whether a token's displayed label predicts the direction of its probability change when the corresponding slider moves, and specificity, whether it predicts change for the token's own attribute rather than a different one. We reuse the setup of \secref{appendix:cos_eval}, namely the same model (\texttt{Qwen3-8B}), decoding parameters (\textit{temperature}${=}1$, \textit{top-p}${=}0$), \texttt{WildChat-WC\textsubscript{Wr}} prompts~\cite{mysorePrototypicalHumanAICollaboration2025a}, and 30 steering settings (10 each for $k{=}1$, $2$, and $3$) with 10 prompts per setting. We take intensity level 5 as the reference condition and label each token \emph{positive} if the attribution ratio $\phi_a(x_i) > 1.1$, \emph{negative} if $\phi_a(x_i) < 0.9$, and \emph{neutral} otherwise, as in the interface.

To obtain the ground-truth direction of change, we fix the sequence generated at the reference level and re-score it under a control vector that changes \emph{only} the target attribute's intensity to a higher (7) or lower (3 and 1) level, computing the per-token change in log-probability
\begin{equation}
\Delta_a(x_i) = \log p_{\boldsymbol{\lambda}'}(x_i \mid x_{<i}, \mathcal{A}) - \log p_{\boldsymbol{\lambda}}(x_i \mid x_{<i}, \mathcal{A}),
\end{equation}
where $\boldsymbol{\lambda}'$ differs from $\boldsymbol{\lambda}$ only in the entry for $a$. Both terms use the full modulated distribution (Equation~3), so the comparison reflects the deployed decoder rather than raw logits. A prediction \emph{agrees} when the sign of $\Delta_a(x_i)$ matches the token's label. We report the \emph{sign-agreement rate} with 95\% bootstrap confidence intervals over settings, using neutral-band tokens as a null control that should sit near chance.\looseness=-1

\namedparagraph{Results}
As summarized in \tabref{tab:attribution}, highlighted tokens agree with the observed direction 81.4\% of the time versus 67.2\% for neutral tokens, a 14.2\,pp gap above the null. The signal is stronger for positive than negative labels (85.1\% vs.\ 58.1\%), consistent with the asymmetry of the leave-one-out ratio, and it degrades only mildly with more simultaneous attributes (83.2\% at $k{=}1$ to 80.8\% at $k{=}3$). It is also attribute-specific: agreement on a token labeled for attribute $a$ is 81.0\% when $a$'s own slider moves but falls to 60.8\% when a different attribute's slider moves, close to the null. These results should be read as token-level rather than sequence-level, since the metric scores probability shifts of the same reference tokens under teacher forcing rather than free re-generation. Overall, they indicate that although the highlights are post-hoc indicators rather than causal explanations, they give a faithful and attribute-specific signal of how a widget reshapes the output distribution.
}

\section{User Study}
% \subsection{Target Outputs}
% \label{appendix:target_versions}
% \todo{...}
\subsection{Participant Information}
\label{appendix:participant_information}
\begin{table*}
  \centering
  \small
  \caption{\formatcaption{Participant information.}{This summary presents participants' demographic information and the self-chosen writing task they explored in the free-form exploration task in our user study. *Writing volume in a typical week; **Prompting experience was self-reported on a 7-point scale, where 1 = Not at all experienced and 7 = Very experienced.}}
  \label{table:demographics}
  \begin{tabular}{llllcl}
    \toprule
    \textbf{ID} & \textbf{Gender} & \textbf{Age} & \textbf{Writing*} & \textbf{Prompting**} & \textbf{Writing Type in Free-Form Exploration} \\ 
    \midrule
    P01 & Male   & 29 & 1--3 hours  & 5 & Research article \\
    P02 & Female & 24 & 8--14 hours & 5 & Homework essay and slide script \\
    P03 & Female & 25 & 15+ hours   & 5 & Activity announcement \\
    P04 & Male   & 24 & 4--7 hours  & 6 & Argumentative essay \\
    P05 & Female & 26 & 4--7 hours  & 5 & Social media post \\
    P06 & Female & 27 & 4--7 hours  & 5 & Email \\
    P07 & Male   & 37 & 15+ hours   & 4 & Research article \\
    P08 & Female & 27 & 4--7 hours  & 5 & Email \\
    P09 & Female & 24 & 4--7 hours  & 5 & Blog post \\
    P10 & Male   & 25 & 1--3 hours  & 5 & Poem \\
    P11 & Male   & 24 & 4--7 hours  & 6 & Reflection after watching a musical \\
    P12 & Female & 28 & 8--14 hours & 6 & Meeting announcement \\
    \bottomrule
  \end{tabular}
\Description{A six-column table summarizing participant demographics and the self-chosen writing task used in the free-form exploration of the user study. The columns are participant ID, gender, age, weekly writing volume, self-reported prompting experience on a 7-point scale, and writing type selected for free-form exploration. The study includes 12 participants, labeled P01 through P12. Ages range from 24 to 37. There are 5 male participants and 7 female participants. Weekly writing volume ranges from 1--3 hours to 15+ hours, with most participants reporting 4--7 hours. Prompting experience ranges from 4 to 6, with most participants reporting 5. The writing tasks selected for free-form exploration are diverse and include research article, homework essay and slide script, activity announcement, argumentative essay, social media post, email, blog post, poem, reflection after watching a musical, and meeting announcement. Research article and email each appear twice, while the other writing types appear once. Overall, the table provides an overview of participant background and the range of writing contexts represented in the study.}
\end{table*}
We recruited 12 participants (7 female, 5 male) aged 24--37 ($M = 26.67$, $SD = 3.68$) via social networks and word of mouth. 
All participants reported using LLM tools, such as ChatGPT, Claude, or Gemini, for writing and editing tasks on a daily basis. 
Regarding their typical weekly writing volume (\eg emails, reports, and posts), two participants reported writing 1--3 hours per week, six reported 4--7 hours, two reported 8--14 hours, and two reported 15+ hours. 
Participants were generally experienced with prompting, with a mean self-rated expertise of $5.17$ ($SD = 0.58$) on a 7-point Likert scale (1 = Not at all, 7 = Very experienced). 
In the free-form exploration, participants used the system for a diverse range of genres, including academic introductions, homework essays, slide scripts, argumentative essays, social media posts, blog posts, and creative writing such as poetry.
\tabref{table:demographics} provides detailed participant information.\looseness=-1

\subsection{Baseline Interface}
\label{appendix:baseline_interface}
\begin{figure*}
  \centering
  \includegraphics[width=0.9\linewidth]{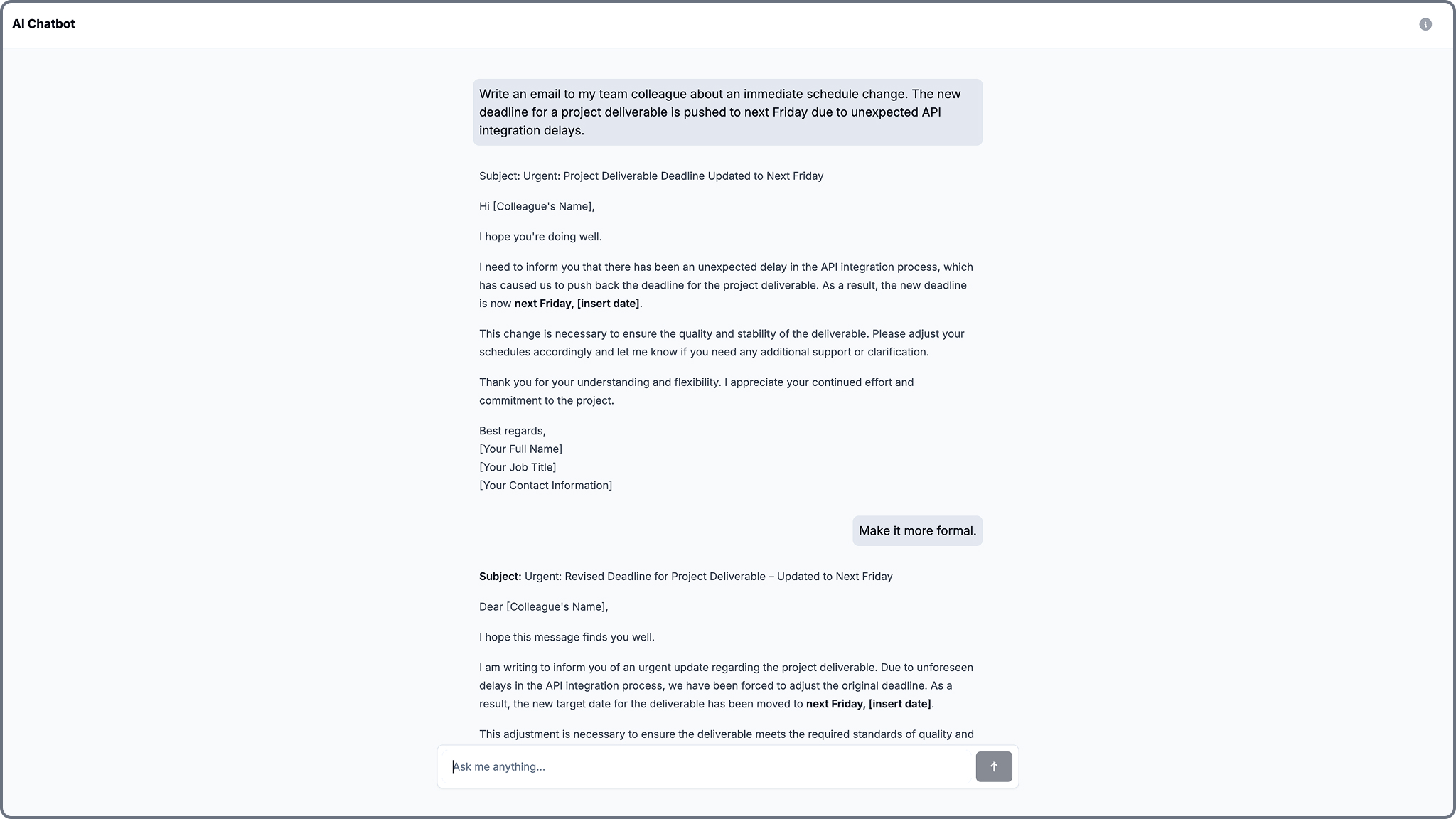}
  \caption{An example screenshot of the baseline interface}
  \Description{An ``AI Chatbot'' interface showing an iterative writing interaction. The interface has a large chat panel with the title ``AI Chatbot'' in the upper left and an information icon in the upper right. In the conversation, the user first enters a request asking the chatbot to write an email to a team colleague about an immediate schedule change, explaining that a project deliverable deadline has been moved to next Friday because of unexpected API integration delays. The chatbot responds with a draft email including the subject line ``Urgent: Project Deliverable Deadline Updated to Next Friday,'' followed by a professional message explaining the delay, the revised deadline, and a request to adjust schedules accordingly. Below that, the user sends a short follow-up message, ``Make it more formal.'' The chatbot then generates a revised version of the email with more formal phrasing, including the updated subject line ``Urgent: Revised Deadline for Project Deliverable – Updated to Next Friday.'' At the bottom of the interface is a text input box labeled ``Ask me anything...'' with a send button. The figure illustrates iterative refinement of writing through natural-language feedback in a chat-based interface.}
  \label{fig:baseline}
  % \vspace{-20pt}
\end{figure*}

The baseline resembles common conversational LLM interfaces such as ChatGPT.
In the baseline, users relied on NL prompting alone to generate outputs through a turn-by-turn conversation.
We used the same underlying model (\texttt{Qwen3-8B}) for output generation as in our system to ensure a fair comparison.
\figref{fig:baseline} shows an example screenshot of the baseline interface.\looseness=-1

\revision{
\subsection{Jeopardy Tasks}
\label{appendix:jeopardy_tasks}
For Jeopardy evaluation, we designed two writing tasks: an email task and an explanation task (\tabref{tab:jeopardy_tasks}).
These tasks reflect a common writing scenario in which we must tailor content to different recipients or audiences~\cite{bellLanguageStyleAudience1984,bitzerRhetoricalSituation1968}.
For each task, we prepared three target versions that differed along multiple preference dimensions based on our taxonomy of the preference space (\tabref{tab:preference_taxonomy}) and prior work~\cite{leeAligningThousandsPreferences2024}, such as audience, tone, depth, and formatting, while preserving the same underlying intent and factual content.
Table~\ref{tab:jeopardy_tasks} summarizes the tasks and target versions.}

\begin{table*}
\footnotesize
\centering
\caption{\formatcaption{Jeopardy evaluation tasks and target preferences.}{Each task includes three target versions that differ along preference dimensions (\eg audience, tone, formality, format, conciseness) based on the taxonomy in \tabref{tab:preference_taxonomy} and prior work~\cite{leeAligningThousandsPreferences2024}, while keeping task facts (\eg dates and key details) constant across versions. Supplemental materials include all target outputs presented to participants.}}
\label{tab:jeopardy_tasks}
\begin{tblr}{
  colspec = {X[1,l] X[2.3,l] X[2.3,l] X[2.3,l] X[2.3,l]},
  hline{1,Z} = {0.8pt, solid},
  row{1} = {font=\bfseries},
  % hlines
}
\textbf{Task} & \textbf{Scenario} & \textbf{Version 1} & \textbf{Version 2} & \textbf{Version 3} \\
\hline
\textbf{Email} &
Write an email about an immediate schedule change for a project deliverable, adapting it to different recipients. &
\textbf{Recipient:} a team colleague \newline
\textbf{Formality:} informal \newline
\textbf{Tone:} informative  \newline
\textbf{Format:} short paragraphs &
\textbf{Recipient:} the manager \newline
\textbf{Formality:} formal \newline
\textbf{Tone:} persuasive\newline
\textbf{Format:} bullet points &
\textbf{Recipient:} the client \newline
\textbf{Formality:} formal \newline
\textbf{Tone:} apologetic \newline
\textbf{Format:} short paragraphs \\
\textbf{Explanation} &
Explain what a large language model is and what it can/cannot do, adapting the explanation to different audiences. &
\textbf{Audience:} tech-savvy users \newline
\textbf{Tone:} professional \newline
\textbf{Conciseness:} detailed \& technical \newline
\textbf{Format:} bullet points &
\textbf{Audience:} general public \newline
\textbf{Tone:} approachable \newline
\textbf{Conciseness:} neutral \newline
\textbf{Format:} short paragraphs &
\textbf{Audience:} middle-schoolers \newline
\textbf{Tone:} playful \& encouraging \newline
\textbf{Conciseness:} simple \& concise \newline
\textbf{Format:} short paragraphs\\
\end{tblr}
\Description{Table summarizing Jeopardy evaluation tasks and preferences of target outputs. It contains two tasks, ``Email'' and ``Explanation,'' and three target versions for each. For the email task, the versions vary by recipient, formality, tone, and format: colleague with informal informative short paragraphs, manager with formal persuasive bullet points, and client with formal apologetic short paragraphs. For the explanation task, the versions vary by audience, tone, conciseness, and format: tech-savvy users with a professional detailed bullet-point explanation, general public with an approachable neutral explanation in short paragraphs, and middle-schoolers with a playful, encouraging, simple explanation in short paragraphs.}
\end{table*}

\subsection{Interview Questions}
\label{appendix:interview_questions}

\begin{enumerate}
\item Walk me through how you approached the task with each interface. What did you do first, and how did you decide what to change next? How did your experience differ
\item How did you decide which parts of the prompt should be made into widgets? Why those?
\item How well did this interface let you steer the output toward what you had in mind? Can you point to a moment when you felt in control, or not in control?
\item How easy was it to express what you wanted (tone, structure, depth, audience)? What felt hard to express or required trial and error?
\item Did the interface match the way you think about your preferences? Were any preferences missing, misrepresented, or hard to specify?
% \item Did you ever try alternatives just to see what would happen? What made you decide to explore vs. stick with one path?
\item Did you realize any new preferences while working? What triggered that realization?
\item When you made a change on the widgets, how easy was it to tell what changed in the output and why it changed? Can you give an example?
% \chao{Did you notice anything you don't want to include from the attribution visualization?}
\item How did you compare different drafts or versions? What cues did you use to decide which version was better or closer to the target?
\item Tell me about a time you felt stuck, confused, or went in circles. What did you do to recover?
\item If you had this interface for your own writing, what tasks or scenarios would you use it for? What would you change to make it fit better?
\end{enumerate}

\revision{
\subsection{Prompt Enrichment Analysis}
Prompt enrichment is designed to address the cold-start problem in preference exploration. Because enrichment changes prompts and can influence generation, it is optional: users may edit or remove it. 
We analyzed how users accepted or rejected the suggested content in enriched prompts as a validation of our prompt enrichment method.
Among 24 first prompts sent to the system, 22 were submitted after enrichment; these had a 93\% word-level acceptance, with 17 used directly, 4 edited, and 1 fully rejected.
}

\end{document}